\documentclass{rspublic}

\usepackage{amsfonts,latexsym,rotate}

\usepackage{QC,royalsociety_cite}

\def\bm#1{\mathchoice{\mbox{\boldmath{$\displaystyle #1$}}}%
{\mbox{\boldmath{$\textstyle #1$}}}%
{\mbox{\boldmath{$\scriptstyle #1$}}}%
{\mbox{\boldmath{$\scriptscriptstyle #1$}}}}

\def\F{{\mathbb{F}}}

\def\ket#1{|#1\rangle}

\def\rev{{\text{\scriptsize rev}}}
\def\trace{\mathop{{\rm tr}}\nolimits}
\newtheorem{defi}[theorem]{Definition}

\title{Cyclic Quantum Error-Correcting Codes and Quantum Shift Registers}

\author[M.~Grassl and Th.~Beth]{Markus Grassl and Thomas Beth}

\affiliation{Institut f{\"u}r Algorithmen und Kognitive Systeme\\
Universit{\"a}t Karlsruhe, Am Fasanengarten 5, 76\,128 Karlsruhe,
Germany.} 

\begin{document}
\label{firstpage}

\maketitle
\begin{abstract}{Quantum error-correcting codes, linear  shift
registers, quantum computing}
We transfer the concept of linear feed-back shift registers to quantum
circuits. It is shown how to use these quantum linear shift registers
for encoding and decoding cyclic quantum error-correcting codes.
\end{abstract}

%==========================================================================
\section{Introduction}
%==========================================================================

Quantum error-correction will be an essential building-block for the
physical implementation of a quantum computer since it is unlikely
that the coherence time of a quantum mechanical system is long enough
to perform any computation of interest, such as factoring large
numbers (see \noparcite{Sho94}). The last years have seen a great
progress in the theory of quantum error-correcting codes (see, e.g.,
\noparcite{KnLa97,CRSS98}). The algorithmic aspect of encoding and
decoding, however, has hardly been addressed, yet.

\cite{ClGo97} gave a general construction for encoding circuits, but
not for decoding. \cite{BeGr98} illustrated how to derive decoding
circuits for quantum error-correcting codes in general. In this paper,
we present a technique for encoding and decoding tailored to cyclic
quantum error-correcting codes. The resulting quantum circuits are
based on the quantum version of linear feed-back shift
registers. Hence, these circuits possess a highly regular structure
and are especially suited for systems with inherent cyclic symmetries,
e.g., circular ion traps. Linear feed-back shift registers fit also to
a heterogeneous system---such as optically trapped atoms combined with
a cavity---where one part of the system---e.g., the cavity---acts as
bus for the feed-back.

The paper is organised as follows: Assuming that the reader is
familiar with the concept of quantum computation in general (see,
e.g., \cite{Ber97,Ste98:QC}), we start with an introduction to
(classical) cyclic error-correcting codes. Then we present linear
shift registers, firstly in their classical, secondly in their quantum
version. In \S\ref{cyclicQECC} quantum circuits for encoding and
decoding cyclic quantum-error correcting codes are presented. We
conclude with an illustrating example and final remarks.

%==========================================================================
\section{Cyclic Codes}
%==========================================================================
In this section we recall some properties of (classical) cyclic
codes. A good reference is, e.g., \cite{MS77}.

%--------------------------------------------------------------------------
\subsection{Polynomial Description}
%--------------------------------------------------------------------------
A cyclic code $C=[N,K]_q$ of length $N$ and dimension $K$ over a
finite field $\F_q=GF(q)$ is a $K$-dimensional subspace of $\F_q^N$
that is invariant under cyclic shifting the coordinates, i.e., for a
codeword $\bm{c}=(c_0,\ldots,c_{N-1})$, the cyclic shift
$(c_{N-1},c_0,\ldots,c_{N-2})$ is again a codeword. To any codeword
$\bm{c}=(c_0,\ldots,c_{N-1})$ we associate the code polynomial
$\bm{c}(X):=c_0+c_1 X+\ldots+c_{N-1} X^{N-1}=\sum_i c_i X^i$. Cyclic
shifting the codeword $\bm{c}$ corresponds to multiplication of the
polynomial $\bm{c}(X)$ by $X$ and reducing it modulo
$X^N-1$. Furthermore, any linear combination of codewords---and thus
code polynomials---is again a codeword. Altogether, the code
corresponds to an ideal in the ring $\F_q[X]/(X^N-1)$. This ideal is
generated by (the residue class of) a polynomial $\bm{g}(X)$ of degree
$N-K$, the {\em generator polynomial} of $C$. Hence, any code
polynomial $\bm{c}(X)$ can be written as
\begin{equation}\label{codepoly1}
\bm{c}(X)=\bm{i}(X) \bm{g}(X) \bmod X^N-1.
\end{equation}
It can be shown that $\bm{g}(X)$ may be chosen as the unique monic
non-zero polynomial of least degree in the code and that $\bm{g}(X)$
divides $X^N-1$, thus $g_0=\bm{g}(0)\ne 0$. The set of code
polynomials is given by
\begin{equation}\label{codepoly2}
\{\bm{c}(X):\bm{c}\in C\}=\{\bm{i}(X)\bm{g}(X)\mid\deg \bm{i}(X)< K\}.
\end{equation}

%--------------------------------------------------------------------------
\subsection{The Dual of  Cyclic Codes}
%--------------------------------------------------------------------------
For a linear block code $C$ of length $N$ over a field $\F_q$,
the dual code $C^\bot$ is given by
$$
C^\bot:=\{ \bm{v}\in\F_q^N\mid\forall\bm{c}\in C:
\bm{c}\cdot\bm{v}=0\}.  
$$
Here $\bm{c}\cdot\bm{v}:=\sum_i c_i v_i$ is the usual inner product of
the vectors $\bm{c}$ and $\bm{v}$.

Obviously, the dual of a cyclic code is cyclic, too. The generator polynomial
$\bm{g}^\bot(X)$ of the dual code is given by
\begin{equation}\label{dualGenPoly}
\bm{g}^\bot(X)=h_0^{-1}\bm{h}^{\rev}(X)\quad
\mbox{where $\bm{g}(X)\bm{h}(X)=X^N-1$.}
\end{equation}
(Note that $h_0\ne 0$ since $\bm{h}(X)|X^N-1$.) Here
$\bm{h}^{\rev}(X)$ denotes the reciprocal polynomial of
$\bm{h}(X)=\sum_i h_i X^i$ obtained by reversing the sequence of
coefficients, i.e., 
\begin{eqnarray*}
\bm{h}^{\rev}(X)&:=&h_0 X^{\deg \bm{h}(X)}+\ldots+h_{\deg \bm{h}(X)} X^0\\
&=&X^{\deg \bm{h}(X)}\bm{h}(1/X).
\end{eqnarray*}

%--------------------------------------------------------------------------
\subsection{The Syndrome of Cyclic Codes}
%--------------------------------------------------------------------------
There are several ways to check whether a given vector $\bm{r}$
resp.{} polynomial $\bm{r}(X)$ is an element of a cyclic code
$C$. From equation~(\ref{codepoly1}), any code polynomial is a
multiple of the generator polynomial $\bm{g}(X)$. Therefore the
syndrome polynomial $\bm{s}(X)$ can be defined as
\begin{equation}\label{polySyndrome}
\bm{s}(X):=\bm{r}(X)\bmod \bm{g}(X).
\end{equation}
The syndrome polynomial is zero if and only if $\bm{r}\in C$, and its
degree is less than $N-K$ otherwise.

Another way to check whether a polynomial $\bm{r}(X)$ belongs to a
code $C$ generated by $\bm{g}(X)$ is the following: Recall that
$\bm{h}(X)=(X^N-1)/\bm{g}(X)$ and that every codeword is a multiple of
$\bm{g}(X)$. Hence $\bm{h}(X)$ can be used as a {\em check polynomial}
with
\begin{equation}\label{checkPoly}
\bm{r}(X)\in C \Longleftrightarrow \bm{r}(X)\bm{h}(X)=\bm{0} \bmod (X^N-1).
\end{equation}

%--------------------------------------------------------------------------
\subsection{Weakly Self-Dual Cyclic Codes}
%--------------------------------------------------------------------------
The construction of quantum error-correcting codes presented in
\S\ref{cyclicQECC} is based on weakly self-dual classical codes, i.e.,
codes $C$ with $C\le C^\bot$. For cyclic codes, a code $C_1$ with
generator polynomial $\bm{g}_1(X)$ is contained in the code $C_2$ with
generator polynomial $\bm{g}_2(X)$ iff $\bm{g}_2(X)$ divides
$\bm{g}_1(X)$. Thus a cyclic code with generator polynomial
$\bm{g}(X)$ is weakly self-dual iff the generator polynomial
$\bm{g}^\bot(X)=h_0^{-1}\bm{h}^{\text{\scriptsize rev}}(X)$ of the
dual code divides $\bm{g}(X)$. In combination with
equation~(\ref{dualGenPoly}) we get the following identities:
\begin{eqnarray}
\bm{g}^\bot(X)&=&h_0^{-1}\bm{h}^{\rev}(X);\nonumber\\
\bm{g}(X)&=&\bm{g}^\bot(X)\tilde{\bm{g}}(X);\label{genPolyFac}\\
X^N-1&=&h_0^{-1}\bm{h}^{\rev}(X)\tilde{\bm{g}}(X)\bm{h}(X).\label{factorization1}
\end{eqnarray}
For a cyclic code $C=[N,K]$ of length $N$ and dimension $K$, the
degrees of the polynomials are as follows:
\begin{eqnarray*}
\deg\bm{g}^\bot(X)&=&K;\\
\deg\bm{g}(X)&=&N-K;\\
\deg\tilde{\bm{g}}(X)&=&N-2K.
\end{eqnarray*}

Next we characterise weakly self-dual cyclic codes in terms of the
factorisation of $X^N-1$ into irreducible polynomials over the field
$\F_q$. As $X^N-1$ is (up to a constant) a self-reciprocal polynomial,
for any factor $\bm{f}(X)$ of $X^N-1$, $\bm{f}^\rev(X)$ is a factor as
well. Hence we can write the factorisation of $X^N-1$ as
$$
X^N-1=\prod_j \bm{r}_j(X) \prod_i \bm{p}_i(X) \prod_i \bm{p}_i^\rev(X)
$$
where the polynomials $\bm{r}_j(X)$ are the (up to a constant)
self-reciprocal factors. From equation~(\ref{factorization1}) follows
that $\bm{h}(X)$ and $\bm{h}^\rev(X)$ have no common factor, hence
each of the self-reciprocal polynomials $\bm{r}_j(X)$ is a factor of
$\tilde{\bm{g}}(X)$, i.e.,
\begin{equation}\label{selfrecFactors}
\prod_j\bm{r}_j(X):=\bm{r}(X)
\qquad\mbox{and}\qquad
\bm{r}(X)|\tilde{\bm{g}}(X).
\end{equation}
Furthermore, for each $i$ at least one of the polynomials
$\bm{p}_i(X)$ and $\bm{p}_i^\rev(X)$ is a factor of $\bm{g}(X)$. 

We conclude this section by a statement about the weights of the
codewords of weakly-self dual cyclic binary codes.
\begin{theorem}
Any weakly self-dual cyclic binary code of odd length is doubly even,
i.e., the weight of any codeword is divisible by four.
\end{theorem}
\begin{proof}
The generator polynomial of the code $C$ can be written as
\begin{equation}\label{sumGenPoly}
\bm{g}(X)=\sum_{i=1}^w X^{d_i}\quad\mbox{where $d_1=0<d_2<\ldots<d_w<N$.}
\end{equation}
The dual code $C^\bot$ has generator polynomial
$\bm{g}^\bot(X)=\bm{h}^\rev(X)$, and its check polynomial is
$(X^N-1)/\bm{h}^\rev(X)=\bm{g}^\rev(X)$. From $C\le C^\bot$ and
equation~(\ref{checkPoly}) we obtain
$$
\bm{f}(X):=\bm{g}(X)\bm{g}^\rev(X)=\bm{a}(X)(X^N-1).
$$
From equation~(\ref{selfrecFactors}) follows that
$\bm{r}(X)^2|\bm{g}(X)\bm{g}^\rev(X)$ and thus $\bm{r}(X)|\bm{a}(X)$,
in particular, $(X+1)|\bm{a}(X)$. Hence the number of terms in
$\bm{a}(X)$ is even. The degree of $\bm{g}(X)$ is less than $N$, and
therefore the degree of $\bm{a}(X)$ is less than $N$, too. This
implies that in the summation $X^N\bm{a}(X)-\bm{a}(X)$ no terms cancel
each other, showing that the number of terms in $\bm{f}(X)$, denoted
by $\#\bm{f}$, is divisible by four.

On the other hand, from equation~(\ref{sumGenPoly}), $\bm{f}(X)$ can be
written as
\begin{equation}\label{summation}
\bm{f}(X)=\sum_{i=1}^w X^{d_i}
  \left(X^{d_w}\sum_{j=1}^w X^{-d_j}\right)=
X^{d_w}\sum_{i,j=1}^w X^{d_i-d_j}.
\end{equation}
Again from equation~(\ref{selfrecFactors}), we conclude that
$(X+1)|\bm{g}(X)$, and thus the number of terms $w$ of $\bm{g}(X)$ is
even. Hence for $i=j$ all terms $X^{d_i-d_j}$ in the summation
(\ref{summation}) cancel each other. For the remaining $w(w-1)$ terms,
two terms cancel each other iff $d_i-d_j=d_k-d_l$. But then we have
also $d_j-d_i=d_l-d_k$, so in total four terms are cancelled. Hence
$\#\bm{f}=w(w-1)-4m$ for some integer $m$. We already know that
$\#\bm{f}$ is divisible by four. Therefore $w(w-1)$ must also be
divisible by four which implies that $w$ is divisible by four since
$w-1$ is odd.

From equation~(\ref{codepoly2}) follows that $\{X^i\bm{g}(X):
i=0,\ldots,N-d_w-1\}$ is a vector space basis of the code. The weight
of each of these vectors is divisible by four. Being a weakly
self-dual code, the inner product of any two codewords is zero, i.e.,
the number of common ones is even. This implies that the weight of the
sum of two codewords which are doubly-even is again divisible by
four. (For the last implication see also \cite[Ch.~1, $\S$8,
Problem~(38)]{MS77}.)
\end{proof}
This theorem shows that all\footnote{\cite{Ste98:efficient} observed
that the dual of some primitive narrow sense BCH codes turn out to be
doubly even. At the CCP workshop at the Isaac Newton Institute,
Cambridge, July 1999, he discussed with us the question when a cyclic
code is doubly even.} quantum error-correcting codes derived from
weakly self-dual cyclic binary codes are well suited for
fault-tolerant quantum computing
(cf.~\noparcite{Got98:faulttolerant}). This is reflected by the fact
that these codes admit the bitwise implementation of the operation
$P=\left(\begin{array}{cc}1&0\\0&i\end{array}\right)$ (see \cite[Lemma
4]{Ste98:efficient}).

%==========================================================================
\section{Cyclic Codes and Linear Shift Registers}
%==========================================================================
The basic operations related to cyclic codes are polynomial
multiplication and division. Both can be done using linear shift
registers.
%--------------------------------------------------------------------------
\subsection{Polynomial Multiplication}
%--------------------------------------------------------------------------
From Horner's rule, the multiplication of a polynomial
$\bm{i}(X)=\sum_{j=0}^\mu i_j X^j$ by the (fixed) polynomial
$\bm{g}(X)=\sum_{j=0}^d g_j X^j$ can be written as
\begin{eqnarray*}
\bm{i}(X)\bm{g}(X)&=&
\biggl(\Bigl((i_{\mu}X+i_{\mu-1})X+\ldots
\Bigr)X+i_0\biggr)\bm{g}(X)\\
&=&
\Bigl(\bigl(i_{\mu}\bm{g}(X)X+i_{\mu-1}\bm{g}(X)\bigr)X+\ldots
\Bigr)X+i_0\bm{g}(X).
\end{eqnarray*}
Feeding the sequence $i_{\mu},i_{\mu-1},\ldots,i_0,0,0,\ldots$
(starting with $i_{\mu}$) into the shift register shown in
figure~\ref{forwardShiftRegister} with the register cells initialised
with zero, it outputs the coefficients of $\bm{i}(X)\bm{g}(X)$,
starting with the coefficient of $X^{d+\mu}$.

\begin{figure}[hbt]
\centerline{\scriptsize\unitlength0.75pt
\begin{picture}(105,80)(-15,0)
\put(20,10){\vector(1,0){30}}
\put(50,10){\makebox(0,0){$\bullet$}}
\put(15,2){\makebox(0,0)[l]{input}}
\put(50,60){\vector(1,0){20}}
\put(70,50){\framebox(20,20){}}
\put(90,10){\line(-1,0){40}}
\put(50,10){\vector(0,1){15}}
\put(50,35){\circle{20}}
\put(50,35){\makebox(0,0){$g_0$}}
\put(50,45){\line(0,1){15}}
\end{picture}%
\begin{picture}(60,80)(30,0)
\put(30,60){\vector(1,0){15}}
\put(45,60){\vector(1,0){25}}
\put(70,50){\framebox(20,20){}}
\put(90,10){\line(-1,0){60}}
\put(50,10){\vector(0,1){15}}
\put(50,10){\makebox(0,0){$\bullet$}}
\put(50,35){\circle{20}}
\put(50,35){\makebox(0,0){$g_1$}}
\put(50,45){\line(0,1){20}}
\put(50,45){\vector(0,1){10}}
\put(50,60){\circle{10}}
\end{picture}%
\begin{picture}(28,80)
\multiput(0,10)(0,50){2}{\line(1,0){10}}
\multiput(12,10)(4,0){3}{\line(1,0){2}}
\multiput(12,35)(4,0){3}{.}
\multiput(12,60)(4,0){3}{\line(1,0){2}}
\multiput(24,10)(0,50){2}{\line(1,0){4}}
\end{picture}%
\begin{picture}(60,80)(30,0)
\put(30,60){\vector(1,0){15}}
\put(45,60){\vector(1,0){25}}
\put(70,50){\framebox(20,20){}}
\put(90,10){\line(-1,0){60}}
\put(50,10){\vector(0,1){15}}
\put(50,10){\makebox(0,0){$\bullet$}}
\put(50,35){\circle{20}}
\put(50,35){\makebox(0,0){$g_{d\mbox{-}1}$}}
\put(50,45){\line(0,1){20}}
\put(50,45){\vector(0,1){10}}
\put(50,60){\circle{10}}
\end{picture}%
\begin{picture}(90,80)(30,0)
\put(30,60){\vector(1,0){15}}
\put(45,60){\vector(1,0){45}}
\put(65,52){\makebox(0,0)[l]{output}}
\put(50,10){\line(-1,0){20}}
\put(50,10){\vector(0,1){15}}
\put(50,35){\circle{20}}
\put(50,35){\makebox(0,0){$g_{d}$}}
\put(50,45){\line(0,1){20}}
\put(50,45){\vector(0,1){10}}
\put(50,60){\circle{10}}
\end{picture}}
\centerline{\scriptsize\unitlength0.75pt
\begin{tabular}[b]{p{6.5em}@{\quad}p{6.5em}@{\quad}p{6.5em}@{}c@{}}
\centering
\begin{picture}(40,50)(0,-10)
\put(0,10){\vector(1,0){10}}
\put(10,0){\framebox(20,20){}}
\put(30,10){\vector(1,0){10}}
\end{picture}
&\centering
\begin{picture}(35,20)
\put(0,20){\vector(1,0){15}}
\put(15,20){\vector(1,0){25}}
\put(20,20){\circle{10}}
\put(20,0){\line(0,1){25}}
\put(20,0){\vector(0,1){15}}
\end{picture}
&\centering
\begin{picture}(40,20)(0,-10)
\put(0,10){\vector(1,0){10}}
\put(20,10){\circle{20}}
\put(20,10){\makebox(0,0){$a$}}
\put(30,10){\vector(1,0){10}}
\end{picture}&\\
\centering
clocked \newline $\F_q$-register-cell &
\centering $\F_q$-adder &
\centering
multiplication \newline by $a\in\F_q$
\end{tabular}}
\bigskip
\caption{Circuit diagram for a linear feed-forward shift register to
multiply the input by $\bm{g}(X)$.\label{forwardShiftRegister}}
\end{figure}

From equation~(\ref{codepoly1}) we see that in order to generate a codeword
of a cyclic code with generating polynomial $\bm{g}(X)$, we just
multiply a polynomial $\bm{i}(X)$ by $\bm{g}(X)$ modulo $X^N-1$. From
equation~(\ref{codepoly2}) follows that the degree of $\bm{i}(X)$ can be
chosen to be less than $K$. Then, reduction modulo $X^N-1$ is not
necessary since the degree of the product is less than $N$. Thus from
the circuit shown in figure~\ref{forwardShiftRegister} we can construct
a circuit with $N$ register cells that computes
$\bm{c}(X)=\bm{i}(X)\bm{g}(X)$ in $K$ steps starting with the
initialisation shown in figure~\ref{encodingShiftRegister}.  

\begin{figure}[htb]
\centerline{\scriptsize\unitlength0.8pt
\begin{picture}(50,80)(40,0)
\put(50,60){\vector(1,0){20}}
\put(70,50){\framebox(20,20){$0$}}
\put(80,75){\makebox(0,0){$0$}}
\put(90,10){\line(-1,0){40}}
\put(50,10){\vector(0,1){15}}
\put(50,35){\circle{20}}
\put(50,35){\makebox(0,0){$g_0$}}
\put(50,45){\line(0,1){15}}
\end{picture}%
\begin{picture}(60,80)(30,0)
\put(30,60){\vector(1,0){15}}
\put(45,60){\vector(1,0){25}}
\put(70,50){\framebox(20,20){$0$}}
\put(80,75){\makebox(0,0){$1$}}
\put(90,10){\line(-1,0){60}}
\put(50,10){\vector(0,1){15}}
\put(50,10){\makebox(0,0){$\bullet$}}
\put(50,35){\circle{20}}
\put(50,35){\makebox(0,0){$g_1$}}
\put(50,45){\line(0,1){20}}
\put(50,45){\vector(0,1){10}}
\put(50,60){\circle{10}}
\end{picture}%
\begin{picture}(28,80)
\multiput(0,10)(0,50){2}{\line(1,0){10}}
\multiput(12,10)(4,0){3}{\line(1,0){2}}
\multiput(12,35)(4,0){3}{.}
\multiput(12,60)(4,0){3}{\line(1,0){2}}
\multiput(24,10)(0,50){2}{\line(1,0){4}}
\end{picture}%
\begin{picture}(60,80)(30,0)
\put(30,60){\vector(1,0){15}}
\put(45,60){\vector(1,0){25}}
\put(70,50){\framebox(20,20){$0$}}
\put(80,75){\makebox(0,0){$d-1$}}
\put(90,10){\line(-1,0){60}}
\put(50,10){\vector(0,1){15}}
\put(50,10){\makebox(0,0){$\bullet$}}
\put(50,35){\circle{20}}
\put(50,35){\makebox(0,0){$g_{d\mbox{-}1}$}}
\put(50,45){\line(0,1){20}}
\put(50,45){\vector(0,1){10}}
\put(50,60){\circle{10}}
\end{picture}%
\begin{picture}(40,80)(30,0)
\put(30,60){\vector(1,0){15}}
\put(45,60){\line(1,0){25}}
\put(70,10){\line(-1,0){50}}
\put(50,10){\vector(0,1){15}}
\put(50,10){\makebox(0,0){$\bullet$}}
\put(50,35){\circle{20}}
\put(50,35){\makebox(0,0){$g_{d}$}}
\put(50,45){\line(0,1){20}}
\put(50,45){\vector(0,1){10}}
\put(50,60){\circle{10}}
\end{picture}%
\begin{picture}(80,80)
\put(0,60){\vector(1,0){50}}
\put(50,50){\framebox(20,20){$i_0$}}
\put(70,60){\line(1,0){10}}
\put(0,10){\line(1,0){80}}
\end{picture}%
\begin{picture}(40,80)
\put(0,60){\vector(1,0){10}}
\put(10,50){\framebox(20,20){$i_1$}}
\put(30,60){\line(1,0){10}}
\put(0,10){\line(1,0){40}}
\end{picture}%
\begin{picture}(28,80)
\multiput(0,10)(0,50){2}{\line(1,0){10}}
\multiput(12,10)(4,0){3}{\line(1,0){2}}
\multiput(12,60)(4,0){3}{\line(1,0){2}}
\multiput(24,10)(0,50){2}{\line(1,0){4}}
\end{picture}%
\begin{picture}(50,80)
\put(0,60){\vector(1,0){10}}
\put(10,50){\framebox(20,20){$i_{K\mbox{-}1}$}}
\put(30,60){\vector(1,0){20}}
\put(0,10){\line(1,0){50}}
\put(50,60){\line(0,-1){50}}
\end{picture}%
}
\caption{Circuit diagram for encoding a cyclic code of length $N$ and
dimension $K$ with generator polynomial
$\bm{g}(X)$.\label{encodingShiftRegister}}
\end{figure}

One single step of the shift register corresponds to the linear
mapping given by
$$
(r'_0,\ldots,r'_{N-1})=(r_0,\ldots,r_{N-1})\cdot E
$$
where
$$
E=\left(\arraycolsep0.5\arraycolsep
\begin{array}{cccccccc}
0  & 1   &  0  & \cdots & \cdots & \cdots &\cdots & 0\\[-0.45em]
\vdots & \ddots & \ddots &\ddots &&&&\vdots\\[-0.6em]
\vdots && \ddots & \ddots &\ddots &&&\vdots\\[-0.6em]
\vdots &&& \ddots & \ddots &\ddots &&\vdots\\[-0.6em] 
\vdots &&&& \ddots & \ddots &\ddots &\vdots\\[-0.6em]
\vdots &&&&& \ddots & \ddots &0 \\
0  & \cdots & \cdots & \cdots & \cdots & \cdots &0 & 1\\
g_0 & g_1 & \ldots & g_d & 0 & \ldots &0 & 0
\end{array}
\right).
$$

The matrix $E$ can be factored into a cyclic shift and adding multiples of
the first element to several others as follows: 
$$
E=\left(\arraycolsep0.5\arraycolsep
\begin{array}{ccccc}
0  & 1   &  0  & \cdots & 0\\[-0.45em]
\vdots & \ddots & \ddots &\ddots &\vdots\\[-0.6em]
\vdots && \ddots & \ddots &0\\
0  & \cdots & \cdots &0 & 1\\
1 &  0 & \cdots &0 & 0
\end{array}
\right)\cdot
\left(\arraycolsep0.5\arraycolsep
\begin{array}{ccccccc}
g_0  & g_1  & \cdots & g_d & 0  & \cdots & 0\\
0 & 1 & 0 & \cdots & \cdots &\cdots &0\\[-0.45em]
\vdots &\ddots & \ddots & \ddots &&& \vdots\\[-0.6em]
\vdots &&\ddots & \ddots & \ddots && \vdots\\[-0.6em]
\vdots &&&\ddots & \ddots & \ddots & \vdots\\[-0.6em]
\vdots &&&&\ddots & \ddots & 0 \\
0 & \cdots &\cdots &\cdots &\cdots &0 &1
\end{array}
\right)
$$
Since the code does not change if we multiply the generator polynomial
by a non-zero constant, we can assume without loss of generality
$g_0=1$ (note that $g_0\ne 0$) thereby simplifying the second factor.

The $K^{\rm th}$ power of $E$ is given by
$$
E^K=\left(\arraycolsep0.5\arraycolsep
\begin{array}{ccccccc}
&&&&1\\[-0.6em]
&&&&&\ddots\\[-0.6em]
&&&&&&1\\
g_0  & g_1  & \cdots & g_d\\
&g_0  & g_1  & \cdots & g_d\\[-0.6em]
&&\ddots & \ddots &&\ddots\\[-0.6em]
&&&g_0  & g_1  & \cdots & g_d
\end{array}
\right)
$$
showing that indeed $(0,\ldots,0,i_0,\ldots,i_{K-1}) E^K=\bm{c}$ with
$\bm{c}(X)=\bm{i}(X)\bm{g}(X)$ and thus $\bm{c}\in C$.

Similarly, it can be shown that for the initialisation
$(j_0,\ldots,j_{d-1},i_0,\ldots,i_{K-1})$, after $K$ steps the state
of the shift register corresponds to
\begin{equation}\label{multCoset}
\bm{p}(X)= \bm{i}(X)\bm{g}(X)+X^K \bm{j}(X)
\end{equation}
where $\bm{j}(X)=j_{d-1}X^{d-1}+\ldots+j_0$.

%--------------------------------------------------------------------------
\subsection{Polynomial Division}
%--------------------------------------------------------------------------
Similar to shift registers for polynomial multiplication, shift
registers can be constructed for polynomial division. The circuit
shown in figure~\ref{backwardShiftRegister} implements a polynomial
division by a monic polynomial $\bm{g}(X)$ of degree $d$. Feeding the
sequence $f_{\mu},f_{\mu-1},\ldots,f_0$ (starting with $f_{\mu}$) into
the shift register shown in figure~\ref{backwardShiftRegister} with
the register cells initialised with zero, it outputs the coefficients
of $\bm{f}(X) \mathop{\rm div} \bm{g}(X)$, starting with the
coefficient of $X^{\mu-d}$. After $\mu+1$ steps, the contents of the
register cells are the coefficients of $\bm{f}(X) \bmod \bm{g}(X)$.

\begin{figure}[hbt]
\centerline{\scriptsize\unitlength0.89pt
\begin{picture}(105,80)(-15,0)
\put(20,60){\vector(1,0){25}}
\put(15,52){\makebox(0,0)[l]{input}}
\put(45,60){\vector(1,0){25}}
\put(70,50){\framebox(20,20){}}
\put(80,75){\makebox(0,0){$0$}}
\put(90,10){\line(-1,0){40}}
\put(50,10){\vector(0,1){15}}
\put(50,35){\circle{20}}
\put(50,35){\makebox(0,0){$g_0$}}
\put(50,45){\line(0,1){20}}
\put(50,45){\vector(0,1){10}}
\put(50,60){\circle{10}}
\end{picture}%
\begin{picture}(60,80)(30,0)
\put(30,60){\vector(1,0){15}}
\put(45,60){\vector(1,0){25}}
\put(70,50){\framebox(20,20){}}
\put(80,75){\makebox(0,0){$1$}}
\put(90,10){\line(-1,0){60}}
\put(50,10){\vector(0,1){15}}
\put(50,10){\makebox(0,0){$\bullet$}}
\put(50,35){\circle{20}}
\put(50,35){\makebox(0,0){$g_1$}}
\put(50,45){\line(0,1){20}}
\put(50,45){\vector(0,1){10}}
\put(50,60){\circle{10}}
\end{picture}%
\begin{picture}(24,80)
\multiput(0,10)(0,50){2}{\line(1,0){10}}
\multiput(12,10)(4,0){3}{\line(1,0){2}}
\multiput(12,35)(4,0){3}{.}
\multiput(12,60)(4,0){3}{\line(1,0){2}}
\multiput(24,10)(0,50){2}{\line(1,0){4}}
\end{picture}%
\begin{picture}(60,80)(30,0)
\put(30,60){\vector(1,0){15}}
\put(45,60){\vector(1,0){25}}
\put(70,50){\framebox(20,20){}}
\put(80,75){\makebox(0,0){$d-1$}}
\put(90,10){\line(-1,0){60}}
\put(50,10){\vector(0,1){15}}
\put(50,10){\makebox(0,0){$\bullet$}}
\put(50,35){\circle{20}}
\put(50,35){\makebox(0,0){$g_{d\mbox{-}1}$}}
\put(50,45){\line(0,1){20}}
\put(50,45){\vector(0,1){10}}
\put(50,60){\circle{10}}
\end{picture}%
\begin{picture}(60,80)(30,0)
\put(30,60){\vector(1,0){50}}
\put(55,52){\makebox(0,0)[l]{output}}
\put(50,60){\makebox(0,0){$\bullet$}}
\put(30,10){\line(1,0){20}}
\put(50,25){\line(0,-1){15}}
\put(50,35){\circle{20}}
\put(50,35){\makebox(0,0){$-1$}}
\put(50,60){\vector(0,-1){15}}
\end{picture}}
\caption{Circuit diagram for a linear feed-back shift register to 
divide the input by the monic polynomial
$\bm{g}(X)$.\label{backwardShiftRegister}}
\end{figure}

To obtain the syndrome of a cyclic code
(cf.~equation~(\ref{polySyndrome})), we have to compute the remainder
of the polynomial $\bm{r}(X)$ modulo $\bm{g}(X)$. Since the degree of
$\bm{r}(X)$ is less than $N$, we can use the circuit shown in
figure~\ref{syndromeShiftRegister} with $N$ register cells initialised
with $(r_0,\ldots,r_{N-1})$. After $N$ steps, the first $d=N-K$
register cells contain the remainder $\bm{r}(X) \bmod \bm{g}(X)$, and
the last $K$ registers contain $\bm{r}(X) \mathbin{\rm div}
\bm{g}(X)$.
\begin{figure}[htb]
\centerline{\scriptsize\unitlength0.89pt
\begin{picture}(50,100)
\put(0,60){\vector(1,0){20}}
\put(0,60){\line(0,1){40}}
\put(20,50){\framebox(20,20){$r_0$}}
\put(40,60){\line(1,0){10}}
\put(0,100){\line(1,0){50}}
\end{picture}%
\begin{picture}(40,100)
\put(0,60){\vector(1,0){10}}
\put(10,50){\framebox(20,20){$r_1$}}
\put(30,60){\line(1,0){10}}
\put(0,100){\line(1,0){40}}
\end{picture}%
\begin{picture}(28,100)
\multiput(0,60)(0,40){2}{\line(1,0){10}}
\multiput(12,60)(4,0){3}{\line(1,0){2}}
\multiput(12,100)(4,0){3}{\line(1,0){2}}
\multiput(24,60)(0,40){2}{\line(1,0){4}}
\end{picture}%
\begin{picture}(50,100)
\put(0,60){\vector(1,0){10}}
\put(10,50){\framebox(20,20){$r_{K\mbox{-}1}$}}
\put(30,60){\line(1,0){20}}
\put(0,100){\line(1,0){50}}
\end{picture}%
\begin{picture}(60,100)(30,0)
\put(45,60){\vector(1,0){25}}
\put(30,60){\vector(1,0){15}}
\put(70,50){\framebox(20,20){$r_K$}}
\put(30,100){\line(1,0){60}}
\put(90,10){\line(-1,0){40}}
\put(50,10){\vector(0,1){15}}
\put(50,35){\circle{20}}
\put(50,35){\makebox(0,0){$g_0$}}
\put(50,45){\line(0,1){20}}
\put(50,45){\vector(0,1){10}}
\put(50,60){\circle{10}}
\end{picture}%
\begin{picture}(60,80)(30,0)
\put(30,60){\vector(1,0){15}}
\put(45,60){\vector(1,0){25}}
\put(70,50){\framebox(20,20){$r_{K\!\mbox{\tiny+}\!1}$}}
\put(30,100){\line(1,0){60}}
\put(90,10){\line(-1,0){60}}
\put(50,10){\vector(0,1){15}}
\put(50,10){\makebox(0,0){$\bullet$}}
\put(50,35){\circle{20}}
\put(50,35){\makebox(0,0){$g_1$}}
\put(50,45){\line(0,1){20}}
\put(50,45){\vector(0,1){10}}
\put(50,60){\circle{10}}
\end{picture}%
\begin{picture}(24,80)
\multiput(0,10)(0,50){2}{\line(1,0){10}}
\put(0,100){\line(1,0){10}}
\multiput(12,10)(4,0){3}{\line(1,0){2}}
\multiput(12,60)(4,0){3}{\line(1,0){2}}
\multiput(12,35)(4,0){3}{.}
\multiput(12,100)(4,0){3}{\line(1,0){2}}
\multiput(24,10)(0,50){2}{\line(1,0){4}}
\end{picture}%
\begin{picture}(60,80)(30,0)
\put(30,60){\vector(1,0){15}}
\put(45,60){\vector(1,0){25}}
\put(70,50){\framebox(20,20){$r_{N\mbox{-}1}$}}
\put(30,100){\line(1,0){60}}
\put(90,10){\line(-1,0){60}}
\put(50,10){\vector(0,1){15}}
\put(50,10){\makebox(0,0){$\bullet$}}
\put(50,35){\circle{20}}
\put(50,35){\makebox(0,0){$g_{d\mbox{-}1}$}}
\put(50,45){\line(0,1){20}}
\put(50,45){\vector(0,1){10}}
\put(50,60){\circle{10}}
\end{picture}%
\begin{picture}(30,100)(30,0)
\put(30,60){\line(1,0){20}}
\put(50,60){\makebox(0,0){$\bullet$}}
\put(50,60){\vector(0,-1){15}}
\put(50,60){\vector(0,1){40}}
\put(30,100){\line(1,0){20}}
\put(30,10){\line(1,0){20}}
\put(50,25){\line(0,-1){15}}
\put(50,35){\circle{20}}
\put(50,35){\makebox(0,0){$-1$}}
\end{picture}%
}
\caption{Circuit diagram for computing the quotient
$\bm{r}(X)\mathbin{\rm div} \bm{g}(X)$ and the remainder
$\bm{r}(X)\bmod\bm{g}(X)$ of the polynomials $\bm{r}(X)$ and
$\bm{g}(X)$ of degree less than $N$ and $d$,
resp.\label{syndromeShiftRegister}}
\end{figure}

The corresponding matrix is given by
$$
S=\left(\arraycolsep0.5\arraycolsep
\begin{array}{ccccc}
0  & 1   &  0  & \cdots & 0\\[-0.45em]
\vdots & \ddots & \ddots &\ddots &\vdots\\[-0.6em]
\vdots && \ddots & \ddots &0\\
0  & \cdots & \cdots &0 & 1\\
1 &  0 & \ldots &0 & 0
\end{array}
\right)\cdot S_2
$$
where
$$
S_2=
\left(\arraycolsep0.5\arraycolsep
\begin{array}{cccccccc}
1 & 0 & \ldots & 0 & -g_0  & -g_1  & \ldots & -g_{d-1} \\[-0.6em]
&\ddots\\[-0.6em]
&&\ddots\\[-0.68em]
&&&\ddots\\[-0.55em]
&&&&\ddots\\[-0.4em]
&&&&&\ddots\\[-0.55em]
&&&&&&\ddots\\
&&&&&&&1
\end{array}
\right).
$$

\begin{figure}[t]
\centerline{\scriptsize\unitlength0.89pt
\begin{picture}(50,80)
\put(0,60){\vector(1,0){20}}
\put(0,60){\line(0,-1){40}}
\put(20,50){\framebox(20,20){$r_0$}}
\put(40,60){\line(1,0){10}}
\put(0,20){\line(1,0){50}}
\end{picture}%
\begin{picture}(40,80)
\put(0,60){\vector(1,0){10}}
\put(10,50){\framebox(20,20){$r_1$}}
\put(30,60){\line(1,0){10}}
\put(0,20){\line(1,0){40}}
\end{picture}%
\begin{picture}(28,80)
\multiput(0,60)(0,-40){2}{\line(1,0){10}}
\multiput(12,60)(4,0){3}{\line(1,0){2}}
\multiput(12,20)(4,0){3}{\line(1,0){2}}
\multiput(24,60)(0,-40){2}{\line(1,0){4}}
\end{picture}%
\begin{picture}(50,80)
\put(0,60){\vector(1,0){10}}
\put(10,50){\framebox(20,20){$r_{\!N\mbox{-}1}$}}
\put(30,60){\line(1,0){20}}
\put(40,60){\makebox(0,0){$\bullet$}}
\put(40,60){\line(0,-1){40}}
\put(0,20){\line(1,0){40}}
\end{picture}%
\begin{picture}(60,80)(30,0)
\put(45,60){\vector(1,0){25}}
\put(30,60){\vector(1,0){15}}
\put(70,50){\framebox(20,20){$0$}}
\put(90,10){\line(-1,0){40}}
\put(50,10){\vector(0,1){15}}
\put(50,35){\circle{20}}
\put(50,35){\makebox(0,0){$g_0$}}
\put(50,45){\line(0,1){20}}
\put(50,45){\vector(0,1){10}}
\put(50,60){\circle{10}}
\end{picture}%
\begin{picture}(60,80)(30,0)
\put(30,60){\vector(1,0){15}}
\put(45,60){\vector(1,0){25}}
\put(70,50){\framebox(20,20){$0$}}
\put(90,10){\line(-1,0){60}}
\put(50,10){\vector(0,1){15}}
\put(50,10){\makebox(0,0){$\bullet$}}
\put(50,35){\circle{20}}
\put(50,35){\makebox(0,0){$g_1$}}
\put(50,45){\line(0,1){20}}
\put(50,45){\vector(0,1){10}}
\put(50,60){\circle{10}}
\end{picture}%
\begin{picture}(24,80)
\multiput(0,10)(0,50){2}{\line(1,0){10}}
\multiput(12,10)(4,0){3}{\line(1,0){2}}
\multiput(12,35)(4,0){3}{.}
\multiput(12,60)(4,0){3}{\line(1,0){2}}
\multiput(24,10)(0,50){2}{\line(1,0){4}}
\end{picture}%
\begin{picture}(60,80)(30,0)
\put(30,60){\vector(1,0){15}}
\put(45,60){\vector(1,0){25}}
\put(70,50){\framebox(20,20){$0$}}
\put(90,10){\line(-1,0){60}}
\put(50,10){\vector(0,1){15}}
\put(50,10){\makebox(0,0){$\bullet$}}
\put(50,35){\circle{20}}
\put(50,35){\makebox(0,0){$g_{d\mbox{-}1}$}}
\put(50,45){\line(0,1){20}}
\put(50,45){\vector(0,1){10}}
\put(50,60){\circle{10}}
\end{picture}%
\begin{picture}(30,80)(30,0)
\put(30,60){\line(1,0){20}}
\put(50,60){\vector(0,-1){15}}
\put(30,10){\line(1,0){20}}
\put(50,25){\line(0,-1){15}}
\put(50,35){\circle{20}}
\put(50,35){\makebox(0,0){$-1$}}
\end{picture}%
}
\caption{Circuit diagram for syndrome computation for a cyclic code of
length $N$ with (monic) generator polynomial $\bm{g}(X)$ of
degree $d$.\label{syndromeShiftRegister2}}
\end{figure}
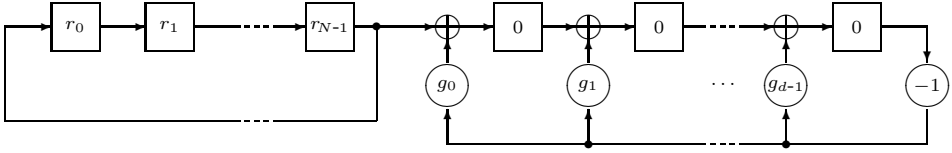

If we are only interested in the remainder and want to keep the
original polynomial $\bm{r}(X)$, a slightly modified version of the
previous circuit can be used (cf.{}
figure~\ref{syndromeShiftRegister2}). After $N$ steps, the first $N$
register cells contain again $\bm{r}(X)$, and the last $d$ register
cells contain $\bm{r}(X)\bmod\bm{g}(X)$. As before, this
transformation can be factored into a shift operation---with two
disjoint cycles of length $N$ and $d$---and a simple linear mapping.

%==========================================================================
\section{Quantum Shift Registers}\label{sec:SelfDualBasis}
%==========================================================================
In this section we show how the linear shift registers presented in
the previous section can be transformed into quantum circuits. For
both linear feed-forward shift registers (for polynomial
multiplication) and linear feed-back shift registers (for polynomial
division) a single basic step can be decomposed into a cyclic shift
followed by a linear mapping of the form
\begin{equation}\label{part2}
\left(\arraycolsep0.5\arraycolsep
\begin{array}{ccccl}
1 & m_2 & m_3 & \ldots & m_{N} \\[-0.3em]
&\ddots\\[-0.6em]
&&\ddots\\[-0.6em]
&&&\ddots\\[-0.6em]
&&&&1
\end{array}
\right).
\end{equation}

First we consider how two implement these mappings for shift registers
over the binary field $\F_2$, then for shift registers over any field
of characteristic two, i.e., over $\F_{2^k}$. In this paper, we
restrict ourselves to fields of characteristic two---corresponding to
qubits---, but the results can easily be generalised to any
characteristic $p>0$.

%--------------------------------------------------------------------------
\subsection{Binary Quantum Shift Registers}
%--------------------------------------------------------------------------
\subsubsection{Cyclic Shifting}
For binary shift register in each cell we have the values zero or
one. Thus we replace each cell by one quantum bit (qubit). The shift
register circuits shown in figures \ref{encodingShiftRegister},
figure~\ref{syndromeShiftRegister}, and \ref{syndromeShiftRegister2} do
all operations in place, i.e., have no input and output. Therefore,
the state of the whole shift register can be represented by $N$
(resp. $N+d$) qubits.

\begin{figure}[hbt]
\centerline{\CNOT(1,2,2)\CNOT(2,1,2)\CNOT(1,2,2)
\quad
\raisebox{17\unitlength}{${}={}$}
\quad
\wires[10](2)%
\begin{picture}(20,40)(0,-10)
\put(0,0){\line(1,1){20}}
\put(0,20){\line(1,-1){20}}
\end{picture}%
\wires[10](2)%
}
\caption{Quantum circuit to swap two qubits.\label{SWAP}}
\end{figure}
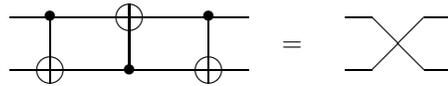

The first part of the basic step of a linear shift register is a
cyclic shift of the qubits. This corresponds to the permutation
$\pi=(1\;\;2\;\ldots\;N)$ which can be written as product of
transpositions
\begin{eqnarray}
(1\;\;2\;\;\ldots\;\;N)&=&(N-1\;\;N)\ldots (2\;\;3)(1\;\;2)\label{cycle1}\\
 &=&(1\;\;N-1)(2\;\;N-2)\ldots(i\;\;N-i)\ldots\nonumber\\
&&\:\cdot(1\;\;N)(2\;\;N-1)\ldots(i\;\;N+1-i)\ldots\label{cycle2}
\end{eqnarray}
(here the leftmost transposition is applied first). While in the first
factorisation there are only transpositions of neighbouring numbers,
the second factorisation is a product of two permutations each of
which is a product of disjoint transpositions. A transposition of two
qubits---a SWAP gate---can be implemented with three controlled not
(CNOT) gates as shown in figure~\ref{SWAP}. (For the graphical
notation of quantum operations see, e.g., \noparcite{BBC95}.)

\begin{figure}[hbt]
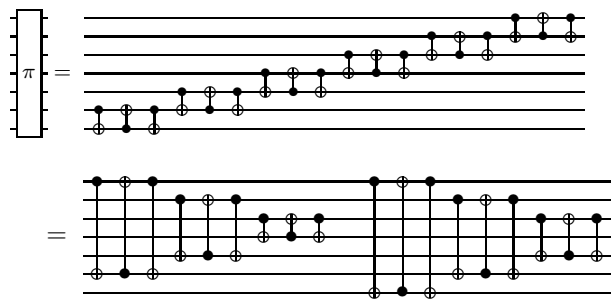

\centerline{\unitlength0.35pt\footnotesize
\multCmultGate[](1,7,7){$\pi$}
  \raisebox{65\unitlength}{${}={}$}
\SWAP(6,7,7)\SWAP(5,6,7)\SWAP(4,5,7)\SWAP(3,4,7)\SWAP(2,3,7)\SWAP(1,2,7)}

\bigskip
\centerline{\unitlength0.35pt\rule{0pt}{0pt}
\hphantom{\multCmultGate[](1,7,7){$\pi$}}
  \raisebox{65\unitlength}{${}={}$}
\SWAP(1,6,7)\SWAP(2,5,7)\SWAP(3,4,7)\wires[30](7)\SWAP(1,7,7)\SWAP(2,6,7)\SWAP(3,5,7)}
\smallskip
\caption{Quantum circuits for cyclic shifting, corresponding to the
different factorisations of the permutation
$\pi=(1\;2\;3\;4\;5\;6\;7)$ given in equation~(\ref{cycle1}) and
equation~(\ref{cycle2}) resp.\label{cycle6}}
\end{figure}

In figure~\ref{cycle6} the circuits corresponding to the
factorisations in equation~(\ref{cycle1}) and equation~(\ref{cycle2})
resp.{} are presented for seven qubits. Both circuits have the same
number of CNOT gates, namely $3(N-1)$, but the second one has only
(constant) depth six if CNOT gates on disjoint sets of qubits can be
performed in parallel.

Note that particular systems may admit simpler implementations of a
single SWAP gate (\noparcite{SKH99}) or the complete cyclic shift.

\subsubsection{Linear Feed-Forward/Feed-Back}
\begin{figure}[hbt]
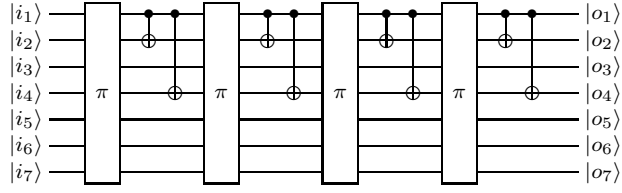

\centerline{\footnotesize\unitlength0.5pt
\inputwires[{$\ket{i_1}$},{$\ket{i_2}$},{$\ket{i_3}$},{$\ket{i_4}$},{$\ket{i_5}$},{$\ket{i_6}$},{$\ket{i_7}$}](7)
\multCmultGate[](1,7,7){$\pi$}\CNOT(1,2,7)\kern-10\unitlength\CNOT(1,4,7)
\multCmultGate[](1,7,7){$\pi$}\CNOT(1,2,7)\kern-10\unitlength\CNOT(1,4,7)
\multCmultGate[](1,7,7){$\pi$}\CNOT(1,2,7)\kern-10\unitlength\CNOT(1,4,7)
\multCmultGate[](1,7,7){$\pi$}\CNOT(1,2,7)\kern-10\unitlength\CNOT(1,4,7)
\outputwires[{$\ket{o_1}$},{$\ket{o_2}$},{$\ket{o_3}$},{$\ket{o_4}$},{$\ket{o_5}$},{$\ket{o_6}$},{$\ket{o_7}$}](7)
}
\smallskip
\caption{Quantum circuit corresponding to $K=4$ steps of a quantum
linear feed-forward shift register for multiplication by the
polynomial $\bm{g}(X)=X^3+X+1$.\label{shiftReg7}}
\end{figure}

The second part of the basic step of a shift register is the linear
transformation given in equation~(\ref{part2}). The first register
cell is unchanged, while multiples of the contents of the first
register cell are added to the other cells. For binary shift
registers, either the value of the first register is added or nothing
is done. The addition of a binary value can be implemented easily, it
corresponds to a CNOT gate. The quantum circuit corresponding to the
linear feed-forward shift register for multiplication by the
polynomial $\bm{g}(X)=X^3+X+1$ is shown in figure~\ref{shiftReg7}. The
shift operation is depicted as a black-box (see
figure~\ref{cycle6}). The two CNOT gates after each shift correspond
to the terms $X^3$ and $X$ in $\bm{g}(X)$.

\begin{figure}[hbt]
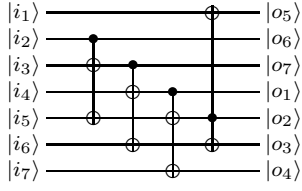

\centerline{\footnotesize\unitlength0.5pt
\inputwires[{$\ket{i_1}$},{$\ket{i_2}$},{$\ket{i_3}$},{$\ket{i_4}$},{$\ket{i_5}$},{$\ket{i_6}$},{$\ket{i_7}$}](7)
\rlap{\CNOT(2,3,7)}\CNOT(2,5,7)
\rlap{\CNOT(3,4,7)}\CNOT(3,6,7)
\rlap{\CNOT(4,5,7)}\CNOT(4,7,7)
\rlap{\CNOT(5,6,7)}\CNOT(5,1,7)
\outputwires[{$\ket{o_5}$},{$\ket{o_6}$},{$\ket{o_7}$},{$\ket{o_1}$},{$\ket{o_2}$},{$\ket{o_3}$},{$\ket{o_4}$}](7)
}
\smallskip
\caption{Alternate version of the quantum circuit shown in figure~\ref{shiftReg7}.\label{shiftReg7simple}}
\end{figure}

An alternate version of this circuit can be obtained if instead of
cyclic shifting the qubits, the other operations are shifted and the
output qubits are re-labelled, as shown in
figure~\ref{shiftReg7simple}. Furthermore, we have combined CNOT gates
with the same control qubit since these gates could be realised with
fewer operations, e.g., in a linear ion trap where the control qubit
is put on the phonon bus.

%--------------------------------------------------------------------------
\subsection{Quantum Shift Registers over Extension Fields}\label{subsec:SelfDualBasis}
%--------------------------------------------------------------------------
\subsubsection{Finite Fields of Characteristic Two}\label{subsubsec:SelfDualBasis}
First, we recall some facts about finite fields (see, e.g.,
\noparcite{Jun93}).

Any finite field $\F_q$ has $q=p^k$ elements where $p$ is a prime
number, the {\em characteristic} of the field. The smallest subset of
$\F_q$ that is a field is called the {\em prime field} of $\F_q$ and
has $p$ elements. Conversely, the field $\F_q$ is an {\em extension
field} of $\F_p$. It can be constructed as $\F_p[X]/(f(X))$ where
$f(X)\in\F_p[x]$ is an irreducible polynomial of degree $k$. The
extension field $\F_q$ is a vector space of dimension $k$ over $\F_p$,
and thus possesses a basis of $k$ linearly independent elements. For a
fixed basis ${\cal B}$, any element of $\F_q$ can be represented by a
vector of length $k$ over $\F_p$. The multiplication by a fixed
element $a\in\F_q$ is a linear mapping and can thus be written as a
$k\times k$ matrix $M_{\cal B}(a)$ over $\F_p$. The trace of $M_{\cal
B}(a)$ is independent of the choice of the basis and defines an
$\F_p$-linear mapping
$$
\trace\colon \F_q \rightarrow \F_p, \quad
 x\mapsto \trace(x):=\trace(M_{\cal B}(x))=\sum_{i=0}^{k-1} x^{p^i}
$$
(for the last equality see, e.g., \cite[Satz 1.24]{Gei94}). 

Finally, we need the definition of the dual basis. Given a basis
${\cal B} = (b_1,\ldots,b_k)$ of a finite field $\F_q$ as
$\F_p$-vector space, the {\em
dual basis} is another basis ${\cal B}^\perp=(b'_1,\ldots,b'_k)$
with
$$
\forall i,j\colon \trace(b_i b'_j)=\delta_{ij}.
$$
Such a dual basis exists for any basis, and the dual basis is unique
(see \cite[Theorem 4.1.1]{Jun93}). A basis that equals its dual
basis is called self-dual.

\subsubsection{Cyclic Shifting}\label{subsubsec:CycShift}
For binary shift register each cell was represented by one
qubit. Fixing a basis ${\cal B}$, each element of the field $\F_{2^k}$
can be represented by a binary vector of length $k$. Hence each cell
of the quantum shift register over the field $\F_{2^k}$ is represented
by $k$ qubits. Cyclic shifting over the extension field is implemented
similarly to the binary case, but now shifting is performed in
parallel in blocks of size $k$. The complexity increases only by the
factor $k$, i.e., shifting can be done with $3k(N-1)$ CNOT gates. The
parallelised version has again constant depth six.

\subsubsection{Linear Feed-Forward/Feed-Back}\label{subsubsec:LinearFeed}
For the second part of the basic step of a shift register we have to
implement the linear transformation given in
equation~(\ref{part2}). Multiples of the contents of the first
register cell are added to the other cells, i.e., we have to implement
the transformations
$$
\ket{x}_1\ket{y}_i \mapsto \ket{x}_1 \ket{m_i x + y}_i
$$
for fixed values $m_i\in\F_{2^k}$. Writing the field elements $x$ and
$y$ as binary vectors of length $k$ with respect to the basis ${\cal
B}=(b_1,\ldots,b_k)$, the multiplication by $m_i$ is a linear
transformation given by the matrix $M:=M_{\cal B}(m_i)$. Now the
transformation can be written as
$$
m_i x + y 
 = \sum_{j=1}^k\left(\sum_{l=1}^k M_{jl} x_l  + y_j\right) b_j
$$
where all operations in parentheses are over the binary field. This
translates directly into a quantum circuit as demonstrated by the
following example.

\begin{figure}[hbt]
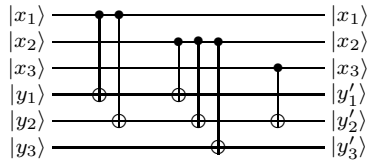

\centerline{\footnotesize\unitlength0.5pt
\inputwires[{$\ket{x_1}$},{$\ket{x_2}$},{$\ket{x_3}$},{$\ket{y_1}$},{$\ket{y_2}$},{$\ket{y_3}$}](6)
\CNOT(1,4,6)\kern-15\unitlength\CNOT(1,5,6) \wires[15](6)
\CNOT(2,4,6)\kern-15\unitlength\CNOT(2,5,6)\kern-15\unitlength\CNOT(2,6,6)
\wires[15](6) \CNOT(3,5,6)
\outputwires[{$\ket{x_1}$},{$\ket{x_2}$},{$\ket{x_3}$},{$|y'_1\rangle$},{$|y'_2\rangle$},{$|y'_3\rangle$}](6)
}
\caption{Quantum circuit implementing the transformation
$\ket{x}\ket{y}\mapsto\ket{x}\ket{\alpha x+y}$.\label{example1}} 
\end{figure}

We consider the field $\F_{2^3}$ with basis ${\cal B}
=(\alpha^3,\alpha^6,\alpha^5)$ where $\alpha^3+\alpha+1=0$. Elements
of $\F_8$ are written as binary column vectors. Multiplication by
$m_2:=\alpha$ corresponds to (left) multiplication of the column
vectors by
\begin{equation}\label{M_alpha}
M_{\cal B}(\alpha)=
\left(\begin{array}{ccc}
1 & 1 & 0 \\
1 & 1 & 1 \\
0 & 1 & 0
\end{array}\right).
\end{equation}
The quantum circuit for the transformation
$\ket{x}\ket{y}\mapsto\ket{x}\ket{\alpha x+y}$ is shown in
figure~\ref{example1}. Conditioned on $x_i$, the $i$th column of
$M_{\cal B}(\alpha)$ is added to the vector $y=(y_1,y_2,y_3)^t$. The
total number of CNOT gates in the circuit is at most
$k+(k-1)^2=k^2-k+1$ since the matrix $M$ is either zero or has full
rank which implies that at most one column (resp.{} row) contains no
zero.

%==========================================================================
\section{Cyclic Quantum Codes}\label{cyclicQECC}
%==========================================================================
%--------------------------------------------------------------------------
\subsection{Binary Codes}
%--------------------------------------------------------------------------
We follow the construction of quantum error-correcting codes from
weakly self-dual binary codes presented by \cite{CaSh96} and
\namecite{Ste96:error}
(\yearcite{Ste96:error},$b$). \nocite{Ste96:multiple} In the
literature, these codes are also referred to as CSS codes.

Given a weakly self-dual linear binary code $C=[N,K]$, the basis
states of the corresponding quantum code are given by
\begin{equation}\label{codeState}
\ket{\psi_j}=\frac{1}{\sqrt{|C|}}
  \sum_{\bm{c}\in C}\ket{\bm{c}+\bm{w}_j}
\end{equation}
where $\{\bm{w}_j:j=1,\ldots, 2^{N-2K}\}$ is a system of
representatives of the cosets $C^\bot/C$. For cyclic
codes, the vector $\bm{c}+\bm{w}_j$ corresponds to the polynomial
$\bm{c}(X)+\bm{w}_j(X)$. Since $\bm{c}\in C$ and
$\bm{w}_j\in C^\bot$, we have
\begin{eqnarray*}
\bm{c}(X)&=&\bm{i}(X)\bm{g}(X)\qquad\mbox{and}\\
\bm{w}_j(X)&=&\bm{j}(X)\bm{g}^\bot(X)
\end{eqnarray*}
for suitably chosen $\bm{i}(X)$ and $\bm{j}(X)$. From
equation~(\ref{genPolyFac}) we get $\bm{g}(X)=\tilde{\bm{g}}(X)\bm{g}^\bot(X)$
and thus
\begin{equation}\label{factoredCodePoly}
\bm{c}(X)+\bm{w}_j(X)
  =\Bigl(\bm{i}(X)\tilde{\bm{g}}(X)+\bm{j}(X)\Bigr)\bm{g}^\bot(X).
\end{equation}
Combining equations~(\ref{codeState}) and (\ref{factoredCodePoly}), we
obtain
\begin{equation}\label{cyclicCodeState}
\ket{\psi_j}=\frac{1}{\sqrt{2^K}}
 \sum_{\deg \bm{i}(X)<K}
  \Bigl|\bigl(\bm{i}(X)\tilde{\bm{g}}(X)+\bm{j}(X)\bigr)\bm{g}^\bot(X)\Bigr\rangle
\end{equation}
where for the polynomial $\bm{f}(X)=f_0+f_1 X+\ldots+f_{N-1}X^{N-1}$,
$\ket{\bm{f}(X)}$ denotes the state
$\ket{f_0}\ket{f_1}\ldots\ket{f_{N-1}}$ .

As $\bm{j}(X)$ is a representative of a coset of the code generated by
$\tilde{\bm{g}}(X)$, without loss of generality we can reduce
$\bm{j}(X)$ modulo $\tilde{\bm{g}}(X)$ and obtain $\deg \bm{j}(X)<\deg
\tilde{\bm{g}}(X)=N-2K$. Hence we get orthogonal basis states
$\ket{\psi_j}$ of the code parameterised by all polynomials
$\bm{j}(X)$ with $\deg\bm{j}(X)<N-2K$. The polynomials
$\bm{i}(X)\tilde{\bm{g}}(X)+\bm{j}(X)$ correspond to elements of the
cosets of the cyclic code $C$. Thus the state $\ket{\psi_j}$ does not
change if we cyclically shift the qubits, i.e., multiply the
polynomial by $X^{N-2K}\bmod (X^N-1)$. Hence
equation~(\ref{cyclicCodeState}) can be written in the (unnormalised)
form
\begin{equation}\label{cyclicCodeState2}
\ket{\psi_j}=
 \sum_{\deg \bm{i}(X)<K}
  \Bigl|\bigl(\bm{i}(X)\tilde{\bm{g}}(X)+X^{N-2K}\bm{j}(X)\bigr)\bm{g}^\bot(X)\Bigr\rangle
\end{equation}
which can be directly translated into an encoding algorithm.

%--------------------------------------------------------------------------
\subsection{Encoding and Decoding}
%--------------------------------------------------------------------------
\subsubsection{Encoding}
\begin{figure}[hbt]
\centerline{\unitlength1.3pt\scriptsize%
\begin{picture}(100,140)(-90,-20)
\multiput(0,0)(0,30){2}{\line(1,0){10}}
\multiput(0,90)(0,30){2}{\line(1,0){10}}
\multiput(0,40)(0,40){2}{\line(1,0){10}}
\multiput(5,11)(0,4){3}{\makebox(0,0){.}}
\multiput(5,56)(0,4){3}{\makebox(0,0){.}}
\multiput(5,101)(0,4){3}{\makebox(0,0){.}}
\multiput(-1,0) (0,30){2}{\makebox(0,0)[r]{$\ket{0}$}}
\multiput(-1,90)(0,30){2}{\makebox(0,0)[r]{$\ket{0}$}}
\multiput(-10,15)(0,90){2}{\makebox(0,0)[r]{$K$ qubits
$\left\{\rule{0pt}{23
\unitlength}\right.$}}
\put(-1,60){\makebox(0,0)[r]{\rotate{$\ket{\bm{j}(X)}$}}}
\put(-10,60){\makebox(0,0)[r]{\normalsize$\ket{\phi_0}\left\{\rule{0pt}{70
\unitlength}\right.$
\scriptsize$N-2K$ qubits
$\left\{\rule{0pt}{25
\unitlength}\right.$}}
\end{picture}%
\begin{picture}(20,140)(0,-20)
\multiput(0,0)(0,30){2}{\line(1,0){6}}
\multiput(6,-4)(0,30){2}{\framebox(8,8){$H$}}
\multiput(10,11)(0,4){3}{\makebox(0,0){.}}
\multiput(14,0)(0,30){2}{\line(1,0){6}}
\multiput(0,40)(0,40){2}{\line(1,0){20}}
\multiput(0,90)(0,30){2}{\line(1,0){20}}
\multiput(20,-5)(0,15){9}{\line(0,1){10}}
\put(20,128){\makebox(0,0)[b]{\normalsize$\ket{\phi_1}$}}
\end{picture}%
\begin{picture}(50,151)(0,-20)
\multiput(0,0)(0,30){2}{\line(1,0){6}}
\put(6,-8){\framebox(28,96){}}
\put(28,85){\line(0,-1){3}}
\put(12,85){\line(0,-1){23}}
\multiput(12,85)(0,-10){3}{\vector(1,0){6.5}}
\multiput(20,85)(0,-10){3}{\circle{3}}
\put(28,85){\line(-1,0){6.5}}
\multiput(29.5,75)(0,-10){2}{\line(-1,0){8}}
\multiput(26,78)(0,-10){2}{\framebox(4,4){}}
\multiput(28,78)(0,-10){2}{\line(0,-1){6}}
\multiput(28,75)(0,-10){2}{\circle{3}}
\multiput(12,50)(0,4){3}{\line(0,1){2}}
\multiput(20,52)(0,4){3}{\makebox(0,0){\tiny.}}
\put(12,48){\line(0,-1){26}}
\multiput(28,50)(0,4){3}{\line(0,1){2}}
\multiput(28,48)(0,-10){2}{\line(0,-1){6}}
\multiput(12,45)(0,-10){2}{\vector(1,0){6.5}}
\multiput(20,45)(0,-10){2}{\circle{3}}
\multiput(29.5,45)(0,-10){2}{\line(-1,0){8}}
\multiput(28,45)(0,-10){2}{\circle{3}}
\multiput(26,38)(0,-10){2}{\framebox(4,4){}}
\multiput(28,38)(0,-10){2}{\line(0,-1){6}}
\multiput(12,10)(0,4){3}{\line(0,1){2}}
\multiput(28,10)(0,4){3}{\line(0,1){2}}
\put(12,-5){\line(1,0){16}}
\put(12,-5){\line(0,1){13}}
\put(28,-5){\line(0,1){3}}
\put(26,-2){\framebox(4,4){}}
\put(28,2){\line(0,1){6}}
\multiput(34,0)(0,30){2}{\line(1,0){16}}
\multiput(0,40)(0,40){2}{\line(1,0){6}}
\multiput(34,40)(0,40){2}{\line(1,0){16}}
\multiput(0,90)(0,30){2}{\line(1,0){50}}
\put(20,-15){\makebox(0,0){$\tilde{\bm{g}}(X)$}}
\multiput(45,-5)(0,15){9}{\line(0,1){10}}
\put(45,128){\makebox(0,0)[b]{\normalsize$\ket{\phi_2}$}}
\end{picture}%
\begin{picture}(80,140)(0,-20)
\multiput(0,0)(0,30){2}{\line(1,0){6}}
\put(6,-8){\framebox(28,136){}}
\put(28,125){\line(0,-1){3}}
\put(12,125){\line(0,-1){13}}
\multiput(12,125)(0,-10){2}{\vector(1,0){6.5}}
\multiput(20,125)(0,-10){2}{\circle{3}}
\put(28,125){\line(-1,0){6.5}}
\put(29.5,115){\line(-1,0){8}}
\put(26,118){\framebox(4,4){}}
\put(28,118){\line(0,-1){6}}
\put(28,115){\circle{3}}
\multiput(12,100)(0,4){3}{\line(0,1){2}}
\multiput(20,101)(0,4){3}{\makebox(0,0){\tiny.}}
\multiput(28,100)(0,4){3}{\line(0,1){2}}
\put(28,98){\line(0,-1){6}}
\put(12,98){\line(0,-1){26}}
\multiput(12,95)(0,-10){2}{\vector(1,0){6.5}}
\multiput(20,95)(0,-10){2}{\circle{3}}
\multiput(29.5,95)(0,-10){2}{\line(-1,0){8}}
\multiput(28,95)(0,-10){2}{\circle{3}}
\multiput(26,88)(0,-10){2}{\framebox(4,4){}}
\multiput(28,88)(0,-10){2}{\line(0,-1){6}}
\multiput(26,38)(0,-10){2}{\framebox(4,4){}}
\multiput(28,48)(0,-10){3}{\line(0,-1){6}}
\put(12,48){\line(0,-1){26}}
\multiput(12,50)(0,4){6}{\line(0,1){2}}
\multiput(28,50)(0,4){6}{\line(0,1){2}}
\multiput(12,10)(0,4){3}{\line(0,1){2}}
\multiput(28,10)(0,4){3}{\line(0,1){2}}
\put(12,-5){\line(1,0){16}}
\put(12,-5){\line(0,1){13}}
\put(28,-5){\line(0,1){3}}
\put(26,-2){\framebox(4,4){}}
\put(28,2){\line(0,1){6}}
\multiput(34,0)(0,30){2}{\line(1,0){16}}
\multiput(0,40)(0,40){2}{\line(1,0){6}}
\multiput(34,40)(0,40){2}{\line(1,0){16}}
\multiput(34,0)(0,30){2}{\line(1,0){16}}
\multiput(0,40)(0,40){2}{\line(1,0){6}}
\multiput(34,40)(0,40){2}{\line(1,0){16}}
\multiput(0,90)(0,30){2}{\line(1,0){6}}
\multiput(34,90)(0,30){2}{\line(1,0){16}}
\put(20,-15){\makebox(0,0){$\bm{g}^\bot(X)$}}
\multiput(45,11)(0,4){3}{\makebox(0,0){.}}
\multiput(45,56)(0,4){3}{\makebox(0,0){.}}
\multiput(45,101)(0,4){3}{\makebox(0,0){.}}
\put(50,60){\makebox(0,0)[l]{\normalsize$\left.\rule{0pt}{70
\unitlength}\right\}\ket{\phi_3}$
}}
\end{picture}%
}
\caption{Quantum circuit for encoding a cyclic quantum-error
correcting code using quantum linear shift registers for
multiplication by $\tilde{\bm{g}}(X)$ and
$\bm{g}^\bot(X)$.\label{QLFSRencoder}}
\end{figure}
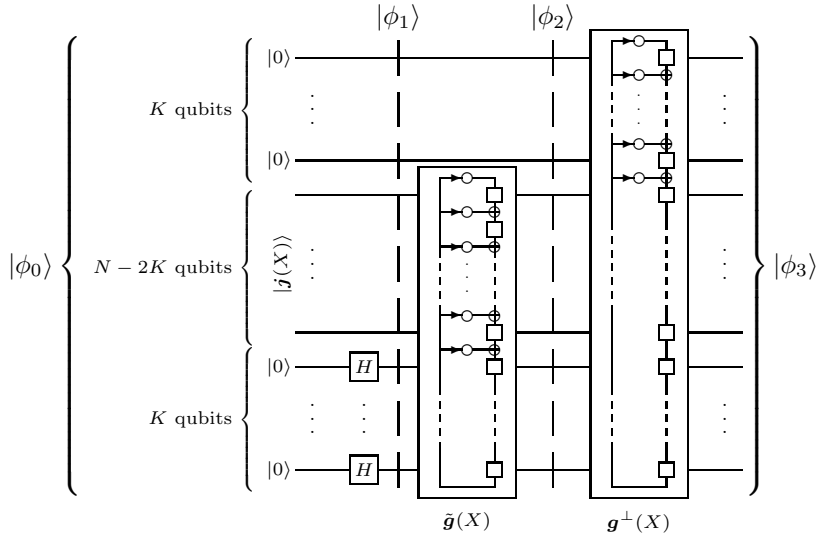

First, we show how to encode quantum information using quantum shift
registers. The initial state of $N-2K$ qubits is embedded into $N$
qubits as follows:
$$
\ket{\phi_0}=\sum_{\deg\bm{j}(X)<N-2K}
\alpha_j\underbrace{\ket{0}\ldots\ket{0}}_K
   \underbrace{\ket{\bm{j}(X)}}_{N-2K}\underbrace{\ket{0}\ldots\ket{0}}_K.
$$
Hadamard transformation of the last $K$ qubits yields the state
$$
\ket{\phi_1}=\sum_{\def\arraystretch{0.5}
\begin{array}{@{}l@{}}
\scriptstyle\deg\bm{i}(X)< K\\
\scriptstyle\deg\bm{j}(X)< N-2K
\end{array}
}\alpha_j\ket{\bm{0}}\ket{\bm{j}(X)}\ket{\bm{i}(X)},
$$
where we have omitted the overall normalisation factor. Using a
quantum linear shift register of length $N-K$ (on the last $N-K$
qubits) for the multiplication by $\tilde{\bm{g}}(X)$, we get (cf.{}
equation~(\ref{multCoset}))
$$
\ket{\phi_2}=\sum_{\def\arraystretch{0.5}
\begin{array}{@{}l@{}}
\scriptstyle\deg\bm{i}(X)< K\\
\scriptstyle\deg\bm{j}(X)< \rlap{$\scriptstyle N-2K$}
\end{array}
}\alpha_j\ket{\bm{0}}\Bigl|\bm{i}(X)\tilde{\bm{g}}(X)+X^{N-2K}\bm{j}(X)\Bigr\rangle.
$$
Finally, in order to multiply by $\bm{g}^\bot(X)$ we use a quantum
shift register of length $N$ and obtain the desired state (cf.{}
equation~(\ref{cyclicCodeState2}))
\begin{equation}\label{phi3}
\ket{\phi_3}=\!\!\sum_{\def\arraystretch{0.5}
\begin{array}{@{}l@{}}
\scriptstyle\deg\bm{i}(X)< \rlap{$\scriptstyle K$}\\
\scriptstyle\deg\bm{j}(X)< \rlap{$\scriptstyle N-2K$}
\end{array}
}\alpha_j\Bigl|\bigl(\bm{i}(X)\tilde{\bm{g}}(X)+X^{N-2K}\bm{j}(X)\bigr)\bm{g}^\bot(X)\Bigr\rangle.
\end{equation}
The whole encoding process is sketched in figure~\ref{QLFSRencoder}.

\subsubsection{Decoding}
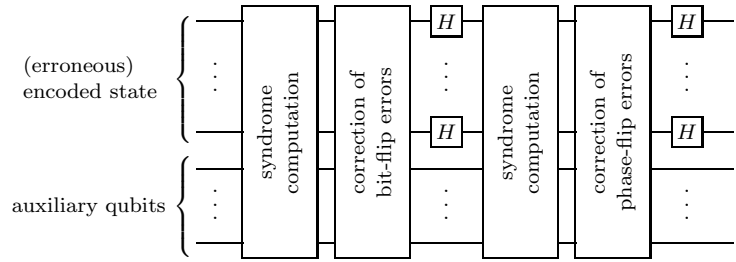
\begin{figure}[hbt]
\centerline{\footnotesize\unitlength0.7pt
\begin{picture}(90,140)(-70,-10)
\multiput(0,0)(0,40){2}{\line(1,0){20}}
\multiput(0,60)(0,60){2}{\line(1,0){20}}
\multiput(10,14)(0,7){3}{\makebox(0,0){.}}
\multiput(10,83)(0,7){3}{\makebox(0,0){.}}
\put(0,90){\makebox(0,0)[r]{\begin{tabular}{l}(erroneous)\\encoded
state\end{tabular}$\left\{\rule{0pt}{40\unitlength}\right.$}}
\put(0,20){\makebox(0,0)[r]{auxiliary qubits
$\left\{\rule{0pt}{30\unitlength}\right.$}}
\end{picture}%
\begin{picture}(50,0)(0,-10)
\multiput(0,0)(0,40){2}{\line(1,0){5}}
\multiput(0,60)(0,60){2}{\line(1,0){5}}
\put(5,-8){\framebox(40,136){\rotate{\begin{tabular}{c}syndrome\\
computation\end{tabular}}}}
\multiput(45,0)(0,40){2}{\line(1,0){5}}
\multiput(45,60)(0,60){2}{\line(1,0){5}}
\end{picture}%
\begin{picture}(50,0)(0,-10)
\multiput(0,0)(0,40){2}{\line(1,0){5}}
\multiput(0,60)(0,60){2}{\line(1,0){5}}
\put(5,-8){\framebox(40,136){\rotate{\begin{tabular}{c}correction of\\
bit-flip errors\end{tabular}}}} 
\multiput(45,0)(0,40){2}{\line(1,0){5}}
\multiput(45,60)(0,60){2}{\line(1,0){5}}
\end{picture}%
\begin{picture}(30,0)(0,-10)
\multiput(0,0)(0,40){2}{\line(1,0){30}}
\put(7,52){\framebox(16,16){$H$}}
\put(7,112){\framebox(16,16){$H$}}
\multiput(0,60)(0,60){2}{\line(1,0){7}}
\multiput(23,60)(0,60){2}{\line(1,0){7}}
\multiput(15,14)(0,7){3}{\makebox(0,0){.}}
\multiput(15,83)(0,7){3}{\makebox(0,0){.}}
\end{picture}%
\begin{picture}(50,0)(0,-10)
\multiput(0,0)(0,40){2}{\line(1,0){5}}
\multiput(0,60)(0,60){2}{\line(1,0){5}}
\put(5,-8){\framebox(40,136){\rotate{\begin{tabular}{c}syndrome\\
computation\end{tabular}}}}
\multiput(45,0)(0,40){2}{\line(1,0){5}}
\multiput(45,60)(0,60){2}{\line(1,0){5}}
\end{picture}%
\begin{picture}(50,0)(0,-10)
\multiput(0,0)(0,40){2}{\line(1,0){5}}
\multiput(0,60)(0,60){2}{\line(1,0){5}}
\put(5,-8){\framebox(40,136){\rotate{\begin{tabular}{c}correction of\\
phase-flip errors\end{tabular}}}} 
\multiput(45,0)(0,40){2}{\line(1,0){5}}
\multiput(45,60)(0,60){2}{\line(1,0){5}}
\end{picture}%
\begin{picture}(30,0)(0,-10)
\multiput(0,0)(0,40){2}{\line(1,0){40}}
\put(7,52){\framebox(16,16){$H$}}
\put(7,112){\framebox(16,16){$H$}}
\multiput(0,60)(0,60){2}{\line(1,0){7}}
\multiput(23,60)(0,60){2}{\line(1,0){17}}
\multiput(15,14)(0,7){3}{\makebox(0,0){.}}
\multiput(15,83)(0,7){3}{\makebox(0,0){.}}
\end{picture}%
}
\caption{General decoding scheme for a quantum error-correcting code
constructed from a weakly self-dual binary code.\label{decScheme}}
\end{figure}

The general outline of the decoding procedure for CSS codes is shown
in figure~\ref{decScheme}. First, errors corresponding to tensor
products of identity and the Pauli matrix $\sigma_x$ (bit-flip errors)
are corrected. Then, a Hadamard transformation interchanges phase-flip
errors (corresponding to $\sigma_z$) with respect to the original
basis and bit-flip errors with respect to the transformed basis. For
quantum error-correcting codes derived from weakly self-dual binary
codes, both steps are essentially the same. Therefore, we describe
only the first step.

\begin{figure}[hbt]
\centerline{\unitlength1.5pt\scriptsize
\begin{picture}(50,140)(-40,-20)
\multiput(0,0)(0,30){2}{\line(1,0){10}}
\put(0,40){\line(1,0){10}}
\multiput(0,60)(0,40){2}{\line(1,0){10}}
\put(0,110){\line(1,0){10}}
\multiput(5,11)(0,4){3}{\makebox(0,0){.}}
\multiput(5,76)(0,4){3}{\makebox(0,0){.}}
\multiput(-1,0)(0,30){2}{\makebox(0,0)[r]{$\ket{0}$}}
\put(-1,40){\makebox(0,0)[r]{$\ket{0}$}}
\put(-10,20){\makebox(0,0)[r]{$K$ qubits
$\left\{\rule{0pt}{28
\unitlength}\right.$}}
\put(-5,85){\makebox(0,0)[r]{\rotate{encoded state}}}
\put(-10,85){\makebox(0,0)[r]{
$N$ qubits
$\left\{\rule{0pt}{30
\unitlength}\right.$}}
\end{picture}%
\begin{picture}(40,140)(0,-20)
\put(6,-8){\framebox(28,126){}}
\put(28,116){\vector(0,-1){4}}
\put(16,116){\line(1,0){12}}
\put(16,116){\line(0,-1){24}}
\multiput(26,108)(0,-10){2}{\framebox(4,4){}}
\multiput(28,108)(0,-10){2}{\vector(0,-1){6}}
\put(26,58){\framebox(4,4){}}
\put(28,58){\vector(0,-1){11.5}}
\put(28,55){\circle*{1}}
\put(28,55){\line(-1,0){12}}
\multiput(16,71)(0,4){5}{\line(0,1){2}}
\multiput(28,71)(0,4){5}{\line(0,1){2}}
\put(28,68){\vector(0,-1){6}}
\put(16,68){\line(0,-1){13}}
\put(28,45){\line(0,-1){3}}
\put(12,45){\line(0,-1){23}}
\multiput(12,45)(0,-10){3}{\vector(1,0){6.5}}
\multiput(20,45)(0,-10){3}{\circle{3}}
\multiput(30,45)(0,-10){3}{\line(-1,0){8.5}}
\multiput(26,38)(0,-10){2}{\framebox(4,4){}}
\multiput(28,48)(0,-10){3}{\line(0,-1){6}}
\multiput(28,45)(0,-10){3}{\circle{3}}
\multiput(12,10)(0,4){3}{\line(0,1){2}}
\multiput(20,11)(0,4){3}{\makebox(0,0){\tiny.}}
\multiput(28,10)(0,4){3}{\line(0,1){2}}
\put(12,8){\line(0,-1){13}}
\put(28,8){\line(0,-1){6}}
\put(12,5){\vector(1,0){6.5}}
\put(18.5,-5){\vector(-1,0){6.5}}
\multiput(20,5)(0,-10){2}{\circle{3}}
\put(30,5){\line(-1,0){8.5}}
\put(28,-5){\line(-1,0){6.5}}
\put(28,5){\circle{3}}
\put(26,-2){\framebox(4,4){}}
\put(28,-2){\line(0,-1){3}}
\multiput(0,0)(0,40){2}{\line(1,0){6}}
\put(0,30){\line(1,0){6}}
\multiput(34,0)(0,40){2}{\line(1,0){6}}
\put(34,30){\line(1,0){6}}
\multiput(0,60)(0,50){2}{\line(1,0){6}}
\put(0,100){\line(1,0){6}}
\multiput(34,60)(0,50){2}{\line(1,0){6}}
\put(34,100){\line(1,0){6}}
\put(20,-15){\makebox(0,0){$\bm{g}^\bot(X)$}}
\end{picture}%
\begin{picture}(25,140)(0,-20)
\multiput(0,0)(0,30){2}{\line(1,0){10}}
\put(0,40){\line(1,0){10}}
\multiput(0,60)(0,40){2}{\line(1,0){10}}
\put(0,110){\line(1,0){10}}
\multiput(5,11)(0,4){3}{\makebox(0,0){.}}
\multiput(5,76)(0,4){3}{\makebox(0,0){.}}
\put(10,20){\makebox(0,0)[l]{$\left.\rule{0pt}{28
\unitlength}\right\}$}}
\put(20,20){\makebox(0,0){\rotate{syndrome of bit-flips}}}
\end{picture}%
}
\caption{Computing the syndrome of a cyclic quantum error-correcting
code using a quantum linear feed-back shift
register.\label{QLFSRdecoder}}
\end{figure}
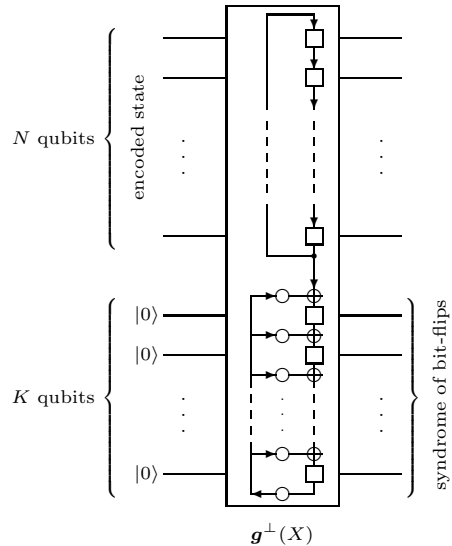

The error-free state (\ref{phi3}) is a superposition of codewords of
the cyclic code generated by $\bm{g}^\bot(X)$. Hence computing the
syndrome $\bm{s}(X)=\bm{r}(X)\bmod \bm{g}^\bot(X)$
(cf. equation~(\ref{polySyndrome})) yields information about the
error. For the computation of the remainder $\bm{r}(X)\bmod
\bm{g}^\bot(X)$, 
we use the quantum version of the linear feed-back shift register
shown in figure~\ref{syndromeShiftRegister2}. The degree of
$\bm{g}^\bot(X)$ is $K$, therefore we need $K$ auxiliary qubits for
the syndrome. The $N$ qubits of the (erroneous) encoded state are
successively fed into the shift register, as depicted in
figure~\ref{QLFSRdecoder}. After $N$ steps, the $K$ auxiliary qubits
contain the syndrome of the bit-flip errors. At this point, a
classical binary syndrome can be obtained by measuring the $K$
syndrome qubits. Then, the corresponding error can be determined using
classical algorithms (e.g., the Berlekamp-Massey algorithm, see
\cite{MS77}). Alternatively, the error may be corrected using quantum
operations that are conditioned on the state of the syndrome qubits.

%--------------------------------------------------------------------------
\subsection{Codes over Fields of Characteristic Two}
%--------------------------------------------------------------------------
\cite{GrGeBe99} showed that CSS codes can also be constructed
using non-binary classical codes. The main idea is to map a code over
an extension field $\F_{2^k}$ to a code over the prime field $\F_2$,
as described in the following definition.
\begin{defi}
Let $C=[N,K]_{2^k}$ be a linear code over the field $\F_{2^k}$ with
basis ${\cal B}=(b_1,\ldots,b_k)$.

Then the {\em binary expansion} of $C$ with respect to the
basis ${\cal B}$ is the linear binary code $C_2=[kN,kK]_2$ given
by
$$C_2:=
\left\{ \left(c_{ij}\right)_{i,j}\in\F_2^{\,kN} \mathrel{\Big|}
  \bm{c}=\left(\textstyle\sum_j c_{ij} b_j\right)_i\in C\right\}.
$$
\end{defi}

The relations between the codes in the previous definition and their
duals are reflected by the following theorem.
\begin{theorem}[see \protect\noparcite{GrGeBe99}]
Let $C=[N,K]_2$ be a linear code over the field $\F_{2^k}$ and
let $C^\bot$ be its dual. Then the dual code of the binary
expansion of $C$ with respect to the basis ${\cal B}$ is the
binary expansion of the dual code $C^\bot$ with respect to the
dual basis ${\cal B}^\bot$, i.e., the following diagram commutes:
$$
\begin{array}{ccc}
C & \longrightarrow & C^{\rlap{$\scriptstyle\bot$}}\\
\left.\llap{\footnotesize basis ${\cal B}$}\rule{0pt}{15pt}\right\downarrow && 
  \left\downarrow\rule{0pt}{15pt}\rlap{\footnotesize dual basis ${\cal B}^\bot$}\right.\\
C_{\rlap{$\scriptstyle2$}} & \longrightarrow & C_{\rlap{$\scriptstyle2$}}^{\rlap{$\scriptstyle\bot$}}\\
\end{array}
$$
\end{theorem}
This theorem shows in particular that the binary code inherits the
property of being weakly self-dual from the code over the extension
field if the binary expansion is with respect to a self-dual basis
(see
\S\ref{sec:SelfDualBasis}$\,\ref{subsec:SelfDualBasis}\,$(\ref{subsubsec:SelfDualBasis})).

If we start with a weakly self-dual cyclic code over the extension
field, the same principles as for cyclic binary codes can be used for
encoding and decoding. We just have to replace the quantum linear
shift registers over the binary field by shift registers over
extension fields (see
\S\ref{sec:SelfDualBasis}$\,\ref{subsec:SelfDualBasis}\,$(\ref{subsubsec:CycShift})
and (\ref{subsubsec:LinearFeed})).

%==========================================================================
\section{Example}
%==========================================================================
To illustrate the preceding, we present quantum circuits based on
quantum shift-registers for quantum Reed-Solomon (${\cal QRS}$) codes
(\noparcite{GrGeBe99}).

We construct a ${\cal QRS}$ code from a Reed-Solomon code
$C=[7,3,5]_8$ over the field $\F_8$. The generator polynomial is
$$
\bm{g}(X)=(X-\alpha^0)(X-\alpha^1)(X-\alpha^2)(X-\alpha^3),
$$
where $\alpha^3+\alpha+1=0$ as above. The dual code $C^\bot=[7,4,4]_8$
is generated by
\begin{eqnarray*}
\bm{g}^\bot(X)&=&(X-\alpha^{-4})(X-\alpha^{-5})(X-\alpha^{-6})\\
&=&(X-\alpha^{3})(X-\alpha^{2})(X-\alpha^{1})\\
&=&\alpha^6(\alpha X^3+X^2+\alpha^2 X+1).
\end{eqnarray*}
Hence $C \le C^\bot$ and $\bm{g}(X)=(X-1)\bm{g}^\bot(X)$.

As self-dual basis of $\F_8$ over $\F_2$ we choose ${\cal
B}=(\alpha^3,\alpha^6,\alpha^5)$. The binary expansions of $C$ and
$C^\bot$ yield binary codes $C_2=[21,9,8]_2$ and
$C_2^\bot=[21,12,5]_2$. Thus the ${\cal QRS}$ code has parameters
${\cal C}=[[21,3,5]]$.

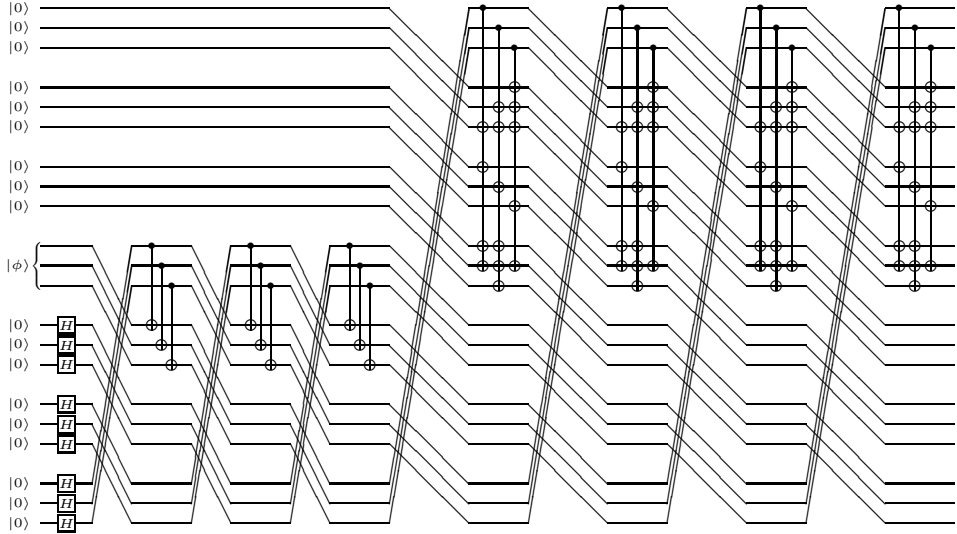
\begin{figure}[hbt]
\centerline{\unitlength0.75pt\tiny%
\begin{picture}(23,260)(-18,0)
\multiput(0,0)(0,40){7}{\line(1,0){5}}
\multiput(0,10)(0,40){7}{\line(1,0){5}}
\multiput(0,20)(0,40){7}{\line(1,0){5}}
\multiput(-5,0)(0,40){3}{\makebox(0,0)[r]{$\ket{0}$}}
\multiput(-5,10)(0,40){3}{\makebox(0,0)[r]{$\ket{0}$}}
\multiput(-5,20)(0,40){3}{\makebox(0,0)[r]{$\ket{0}$}}
\put(-5,130){\makebox(0,0)[r]{$\ket{\phi}$\rlap{$\!\left.\rule{0pt}{13\unitlength}\right\{$}}}
\multiput(-5,160)(0,40){3}{\makebox(0,0)[r]{$\ket{0}$}}
\multiput(-5,170)(0,40){3}{\makebox(0,0)[r]{$\ket{0}$}}
\multiput(-5,180)(0,40){3}{\makebox(0,0)[r]{$\ket{0}$}}
\end{picture}%
%Hadamard transform
\begin{picture}(16,260)
\multiput(0,0)(0,40){3}{\line(1,0){4}}
\multiput(0,10)(0,40){3}{\line(1,0){4}}
\multiput(0,20)(0,40){3}{\line(1,0){4}}
\multiput(4,-4)(0,40){3}{\framebox(8,8){$H$}}
\multiput(4,6)(0,40){3}{\framebox(8,8){$H$}}
\multiput(4,16)(0,40){3}{\framebox(8,8){$H$}}
\multiput(12,0)(0,40){3}{\line(1,0){4}}
\multiput(12,10)(0,40){3}{\line(1,0){4}}
\multiput(12,20)(0,40){3}{\line(1,0){4}}
\multiput(0,120)(0,40){4}{\line(1,0){16}}
\multiput(0,130)(0,40){4}{\line(1,0){16}}
\multiput(0,140)(0,40){4}{\line(1,0){16}}
\end{picture}%
% first shift register
\begin{picture}(30,260)
\multiput(0,0)(0,40){4}{\line(1,0){5}}
\multiput(0,10)(0,40){4}{\line(1,0){5}}
\multiput(0,20)(0,40){4}{\line(1,0){5}}
\multiput(5,0)(0,10){3}{\line(1,6){20}}
\multiput(5,40)(0,40){3}{\line(1,-2){20}}
\multiput(5,50)(0,40){3}{\line(1,-2){20}}
\multiput(5,60)(0,40){3}{\line(1,-2){20}}
\multiput(25,0)(0,40){4}{\line(1,0){5}}
\multiput(25,10)(0,40){4}{\line(1,0){5}}
\multiput(25,20)(0,40){4}{\line(1,0){5}}
\multiput(0,160)(0,40){3}{\line(1,0){30}}
\multiput(0,170)(0,40){3}{\line(1,0){30}}
\multiput(0,180)(0,40){3}{\line(1,0){30}}
\end{picture}%
\begin{picture}(20,260)
\multiput(5,140)(5,-10){3}{\makebox(0,0){$\bullet$}}
\multiput(5,140)(5,-10){3}{\line(0,-1){42.5}}
\multiput(5,100)(5,-10){3}{\circle{5}}
\multiput(0,0)(0,40){7}{\line(1,0){20}}
\multiput(0,10)(0,40){7}{\line(1,0){20}}
\multiput(0,20)(0,40){7}{\line(1,0){20}}
\end{picture}%
\begin{picture}(30,260)
\multiput(0,0)(0,40){4}{\line(1,0){5}}
\multiput(0,10)(0,40){4}{\line(1,0){5}}
\multiput(0,20)(0,40){4}{\line(1,0){5}}
\multiput(5,0)(0,10){3}{\line(1,6){20}}
\multiput(5,40)(0,40){3}{\line(1,-2){20}}
\multiput(5,50)(0,40){3}{\line(1,-2){20}}
\multiput(5,60)(0,40){3}{\line(1,-2){20}}
\multiput(25,0)(0,40){4}{\line(1,0){5}}
\multiput(25,10)(0,40){4}{\line(1,0){5}}
\multiput(25,20)(0,40){4}{\line(1,0){5}}
\multiput(0,160)(0,40){3}{\line(1,0){30}}
\multiput(0,170)(0,40){3}{\line(1,0){30}}
\multiput(0,180)(0,40){3}{\line(1,0){30}}
\end{picture}%
\begin{picture}(20,260)
\multiput(5,140)(5,-10){3}{\makebox(0,0){$\bullet$}}
\multiput(5,140)(5,-10){3}{\line(0,-1){42.5}}
\multiput(5,100)(5,-10){3}{\circle{5}}
\multiput(0,0)(0,40){7}{\line(1,0){20}}
\multiput(0,10)(0,40){7}{\line(1,0){20}}
\multiput(0,20)(0,40){7}{\line(1,0){20}}
\end{picture}%
\begin{picture}(30,260)
\multiput(0,0)(0,40){4}{\line(1,0){5}}
\multiput(0,10)(0,40){4}{\line(1,0){5}}
\multiput(0,20)(0,40){4}{\line(1,0){5}}
\multiput(5,0)(0,10){3}{\line(1,6){20}}
\multiput(5,40)(0,40){3}{\line(1,-2){20}}
\multiput(5,50)(0,40){3}{\line(1,-2){20}}
\multiput(5,60)(0,40){3}{\line(1,-2){20}}
\multiput(25,0)(0,40){4}{\line(1,0){5}}
\multiput(25,10)(0,40){4}{\line(1,0){5}}
\multiput(25,20)(0,40){4}{\line(1,0){5}}
\multiput(0,160)(0,40){3}{\line(1,0){30}}
\multiput(0,170)(0,40){3}{\line(1,0){30}}
\multiput(0,180)(0,40){3}{\line(1,0){30}}
\end{picture}%
\begin{picture}(20,260)
\multiput(5,140)(5,-10){3}{\makebox(0,0){$\bullet$}}
\multiput(5,140)(5,-10){3}{\line(0,-1){42.5}}
\multiput(5,100)(5,-10){3}{\circle{5}}
\multiput(0,0)(0,40){7}{\line(1,0){20}}
\multiput(0,10)(0,40){7}{\line(1,0){20}}
\multiput(0,20)(0,40){7}{\line(1,0){20}}
\end{picture}%
% second shift register
\begin{picture}(50,260)
\multiput(0,0)(0,40){7}{\line(1,0){5}}
\multiput(0,10)(0,40){7}{\line(1,0){5}}
\multiput(0,20)(0,40){7}{\line(1,0){5}}
\multiput(5,0)(0,10){3}{\line(1,6){40}}
\multiput(5,40)(0,40){6}{\line(1,-1){40}}
\multiput(5,50)(0,40){6}{\line(1,-1){40}}
\multiput(5,60)(0,40){6}{\line(1,-1){40}}
\multiput(45,0)(0,40){7}{\line(1,0){5}}
\multiput(45,10)(0,40){7}{\line(1,0){5}}
\multiput(45,20)(0,40){7}{\line(1,0){5}}
\end{picture}%
\begin{picture}(20,260)
\multiput(2,260)(8,-10){3}{\makebox(0,0){$\bullet$}}
\put(2,260){\line(0,-1){132.5}}
\put(10,250){\line(0,-1){132.5}}
\put(18,240){\line(0,-1){112.5}}
% alpha^2
\put(2,200){\circle{5}}
\multiput(10,200)(0,10){2}{\circle{5}}
\multiput(18,200)(0,10){3}{\circle{5}}
% 1
\multiput(2,180)(8,-10){3}{\circle{5}}
% alpha
\multiput(2,140)(0,-10){2}{\circle{5}}
\multiput(10,140)(0,-10){3}{\circle{5}}
\put(18,130){\circle{5}}
\multiput(0,0)(0,40){7}{\line(1,0){20}}
\multiput(0,10)(0,40){7}{\line(1,0){20}}
\multiput(0,20)(0,40){7}{\line(1,0){20}}
\end{picture}%
\begin{picture}(50,260)
\multiput(0,0)(0,40){7}{\line(1,0){5}}
\multiput(0,10)(0,40){7}{\line(1,0){5}}
\multiput(0,20)(0,40){7}{\line(1,0){5}}
\multiput(5,0)(0,10){3}{\line(1,6){40}}
\multiput(5,40)(0,40){6}{\line(1,-1){40}}
\multiput(5,50)(0,40){6}{\line(1,-1){40}}
\multiput(5,60)(0,40){6}{\line(1,-1){40}}
\multiput(45,0)(0,40){7}{\line(1,0){5}}
\multiput(45,10)(0,40){7}{\line(1,0){5}}
\multiput(45,20)(0,40){7}{\line(1,0){5}}
\end{picture}%
\begin{picture}(20,260)
\multiput(2,260)(8,-10){3}{\makebox(0,0){$\bullet$}}
\put(2,260){\line(0,-1){132.5}}
\put(10,250){\line(0,-1){132.5}}
\put(18,240){\line(0,-1){112.5}}
% alpha^2
\put(2,200){\circle{5}}
\multiput(10,200)(0,10){2}{\circle{5}}
\multiput(18,200)(0,10){3}{\circle{5}}
% 1
\multiput(2,180)(8,-10){3}{\circle{5}}
% alpha
\multiput(2,140)(0,-10){2}{\circle{5}}
\multiput(10,140)(0,-10){3}{\circle{5}}
\put(18,130){\circle{5}}
\multiput(0,0)(0,40){7}{\line(1,0){20}}
\multiput(0,10)(0,40){7}{\line(1,0){20}}
\multiput(0,20)(0,40){7}{\line(1,0){20}}
\end{picture}%
\begin{picture}(50,260)
\multiput(0,0)(0,40){7}{\line(1,0){5}}
\multiput(0,10)(0,40){7}{\line(1,0){5}}
\multiput(0,20)(0,40){7}{\line(1,0){5}}
\multiput(5,0)(0,10){3}{\line(1,6){40}}
\multiput(5,40)(0,40){6}{\line(1,-1){40}}
\multiput(5,50)(0,40){6}{\line(1,-1){40}}
\multiput(5,60)(0,40){6}{\line(1,-1){40}}
\multiput(45,0)(0,40){7}{\line(1,0){5}}
\multiput(45,10)(0,40){7}{\line(1,0){5}}
\multiput(45,20)(0,40){7}{\line(1,0){5}}
\end{picture}%
\begin{picture}(20,260)
\multiput(2,260)(8,-10){3}{\makebox(0,0){$\bullet$}}
\put(2,260){\line(0,-1){132.5}}
\put(10,250){\line(0,-1){132.5}}
\put(18,240){\line(0,-1){112.5}}
% alpha^2
\put(2,200){\circle{5}}
\multiput(10,200)(0,10){2}{\circle{5}}
\multiput(18,200)(0,10){3}{\circle{5}}
% 1
\multiput(2,180)(8,-10){3}{\circle{5}}
% alpha
\multiput(2,140)(0,-10){2}{\circle{5}}
\multiput(10,140)(0,-10){3}{\circle{5}}
\put(18,130){\circle{5}}
\multiput(0,0)(0,40){7}{\line(1,0){20}}
\multiput(0,10)(0,40){7}{\line(1,0){20}}
\multiput(0,20)(0,40){7}{\line(1,0){20}}
\end{picture}%
\begin{picture}(50,260)
\multiput(0,0)(0,40){7}{\line(1,0){5}}
\multiput(0,10)(0,40){7}{\line(1,0){5}}
\multiput(0,20)(0,40){7}{\line(1,0){5}}
\multiput(5,0)(0,10){3}{\line(1,6){40}}
\multiput(5,40)(0,40){6}{\line(1,-1){40}}
\multiput(5,50)(0,40){6}{\line(1,-1){40}}
\multiput(5,60)(0,40){6}{\line(1,-1){40}}
\multiput(45,0)(0,40){7}{\line(1,0){5}}
\multiput(45,10)(0,40){7}{\line(1,0){5}}
\multiput(45,20)(0,40){7}{\line(1,0){5}}
\end{picture}%
\begin{picture}(30,260)
\multiput(2,260)(8,-10){3}{\makebox(0,0){$\bullet$}}
\put(2,260){\line(0,-1){132.5}}
\put(10,250){\line(0,-1){132.5}}
\put(18,240){\line(0,-1){112.5}}
% alpha^2
\put(2,200){\circle{5}}
\multiput(10,200)(0,10){2}{\circle{5}}
\multiput(18,200)(0,10){3}{\circle{5}}
% 1
\multiput(2,180)(8,-10){3}{\circle{5}}
% alpha
\multiput(2,140)(0,-10){2}{\circle{5}}
\multiput(10,140)(0,-10){3}{\circle{5}}
\put(18,130){\circle{5}}
\multiput(0,0)(0,40){7}{\line(1,0){30}}
\multiput(0,10)(0,40){7}{\line(1,0){30}}
\multiput(0,20)(0,40){7}{\line(1,0){30}}
\end{picture}%
}
\smallskip
\caption{Encoder for the quantum Reed-Solomon code $[[21,3,5]]$ using
quantum shift registers for the multiplication by
$\tilde{\bm{g}}(X)=X+1$ and $\bm{g}^\bot=\alpha X^3+X^2+\alpha^2
X+1$.\label{QRSencoder}}
\end{figure}

The encoding circuit shown in figure~\ref{QRSencoder} has the same
structure as that in figure~\ref{QLFSRencoder}. First, the $3$-qubit
state (`{\em q-octet}') $\ket{\phi}$ is embedded into $21$ qubits (or
$7$ q-octets) forming the state $\ket{\phi_0}$. Next, three steps of
the quantum shift register for the multiplication by
$\tilde{\bm{g}}(X)=X+1$ follow. In figure~\ref{QRSencoder}, the shift
operation is depicted by a permutation of the lines representing the
qubits. Finally, we have four steps of the quantum shift register for
the multiplication by $\bm{g}^\bot(X)$. We make the normalisation
$g_0=1$ and obtain $\bm{g}^\bot(X)=\alpha X^3+X^2+\alpha^2 X+1$. The
matrices corresponding the multiplication by the non-trivial
coefficients of $\bm{g}^\bot(X)$ are given by
$$
M_{\cal B}(\alpha)=
\left(\begin{array}{ccc}
1 & 1 & 0 \\
1 & 1 & 1 \\
0 & 1 & 0
\end{array}\right)
\qquad\mbox{and}\qquad
M_{\cal B}(\alpha^2)=
\left(\begin{array}{ccc}
0 & 0 & 1 \\
0 & 1 & 1 \\
1 & 1 & 1
\end{array}\right).
$$
As in equation~(\ref{M_alpha}) and figure~\ref{example1}, the structure
of this matrices is reflected by the quantum circuit.

\begin{figure}[hbt]
\centerline{\unitlength0.75pt\tiny%
\begin{picture}(23,260)(-18,0)
\multiput(0,0)(0,40){7}{\line(1,0){5}}
\multiput(0,10)(0,40){7}{\line(1,0){5}}
\multiput(0,20)(0,40){7}{\line(1,0){5}}
\multiput(-5,0)(0,40){3}{\makebox(0,0)[r]{$\ket{0}$}}
\multiput(-5,10)(0,40){3}{\makebox(0,0)[r]{$\ket{0}$}}
\multiput(-5,20)(0,40){3}{\makebox(0,0)[r]{$\ket{0}$}}
\put(-5,130){\makebox(0,0)[r]{$\ket{\phi}$\rlap{$\!\left.\rule{0pt}{13\unitlength}\right\{$}}}
\multiput(-5,160)(0,40){3}{\makebox(0,0)[r]{$\ket{0}$}}
\multiput(-5,170)(0,40){3}{\makebox(0,0)[r]{$\ket{0}$}}
\multiput(-5,180)(0,40){3}{\makebox(0,0)[r]{$\ket{0}$}}
\end{picture}%
% Hadamard transform
\begin{picture}(26,260)
\multiput(0,160)(0,40){3}{\line(1,0){4}}
\multiput(0,170)(0,40){3}{\line(1,0){4}}
\multiput(0,180)(0,40){3}{\line(1,0){4}}
\multiput(4,156)(0,40){3}{\framebox(8,8){$H$}}
\multiput(4,166)(0,40){3}{\framebox(8,8){$H$}}
\multiput(4,176)(0,40){3}{\framebox(8,8){$H$}}
\multiput(12,160)(0,40){3}{\line(1,0){14}}
\multiput(12,170)(0,40){3}{\line(1,0){14}}
\multiput(12,180)(0,40){3}{\line(1,0){14}}
\multiput(0,0)(0,40){4}{\line(1,0){26}}
\multiput(0,10)(0,40){4}{\line(1,0){26}}
\multiput(0,20)(0,40){4}{\line(1,0){26}}
\end{picture}%
% first shift register
\begin{picture}(40,260)
\multiput(5,180)(10,-10){3}{\makebox(0,0){$\bullet$}}
\multiput(5,180)(10,-10){3}{\line(0,-1){42.5}}
\multiput(5,140)(10,-10){3}{\circle{5}}
\multiput(0,0)(0,40){7}{\line(1,0){40}}
\multiput(0,10)(0,40){7}{\line(1,0){40}}
\multiput(0,20)(0,40){7}{\line(1,0){40}}
\end{picture}%
\begin{picture}(40,260)
\multiput(5,220)(10,-10){3}{\makebox(0,0){$\bullet$}}
\multiput(5,220)(10,-10){3}{\line(0,-1){42.5}}
\multiput(5,180)(10,-10){3}{\circle{5}}
\multiput(0,0)(0,40){7}{\line(1,0){40}}
\multiput(0,10)(0,40){7}{\line(1,0){40}}
\multiput(0,20)(0,40){7}{\line(1,0){40}}
\end{picture}%
\begin{picture}(40,260)
\multiput(5,260)(10,-10){3}{\makebox(0,0){$\bullet$}}
\multiput(5,260)(10,-10){3}{\line(0,-1){42.5}}
\multiput(5,220)(10,-10){3}{\circle{5}}
\multiput(0,0)(0,40){7}{\line(1,0){40}}
\multiput(0,10)(0,40){7}{\line(1,0){40}}
\multiput(0,20)(0,40){7}{\line(1,0){40}}
\end{picture}%
% second shift register
\begin{picture}(40,260)
\multiput(0,0)(0,40){7}{\line(1,0){30}}
\multiput(0,10)(0,40){7}{\line(1,0){30}}
\multiput(0,20)(0,40){7}{\line(1,0){30}}
\multiput(5,140)(10,-10){3}{\makebox(0,0){$\bullet$}}
\put(5,140){\line(0,-1){132.5}}
\put(15,130){\line(0,-1){132.5}}
\put(25,120){\line(0,-1){112.5}}
% alpha^2
\put(5,80){\circle{5}}
\multiput(15,80)(0,10){2}{\circle{5}}
\multiput(25,80)(0,10){3}{\circle{5}}
% 1
\multiput(5,60)(10,-10){3}{\circle{5}}
% alpha
\multiput(5,20)(0,-10){2}{\circle{5}}
\multiput(15,20)(0,-10){3}{\circle{5}}
\put(25,10){\circle{5}}
\multiput(0,0)(0,40){7}{\line(1,0){40}}
\multiput(0,10)(0,40){7}{\line(1,0){40}}
\multiput(0,20)(0,40){7}{\line(1,0){40}}
\end{picture}%
\begin{picture}(40,260)
\multiput(5,180)(10,-10){3}{\makebox(0,0){$\bullet$}}
\put(5,180){\line(0,-1){132.5}}
\put(15,170){\line(0,-1){132.5}}
\put(25,160){\line(0,-1){112.5}}
% alpha^2
\put(5,120){\circle{5}}
\multiput(15,120)(0,10){2}{\circle{5}}
\multiput(25,120)(0,10){3}{\circle{5}}
% 1
\multiput(5,100)(10,-10){3}{\circle{5}}
% alpha
\multiput(5,60)(0,-10){2}{\circle{5}}
\multiput(15,60)(0,-10){3}{\circle{5}}
\put(25,50){\circle{5}}
\multiput(0,0)(0,40){7}{\line(1,0){40}}
\multiput(0,10)(0,40){7}{\line(1,0){40}}
\multiput(0,20)(0,40){7}{\line(1,0){40}}
\end{picture}%
\begin{picture}(40,260)
\multiput(5,220)(10,-10){3}{\makebox(0,0){$\bullet$}}
\put(5,220){\line(0,-1){132.5}}
\put(15,210){\line(0,-1){132.5}}
\put(25,200){\line(0,-1){112.5}}
% alpha^2
\put(5,160){\circle{5}}
\multiput(15,160)(0,10){2}{\circle{5}}
\multiput(25,160)(0,10){3}{\circle{5}}
% 1
\multiput(5,140)(10,-10){3}{\circle{5}}
% alpha
\multiput(5,100)(0,-10){2}{\circle{5}}
\multiput(15,100)(0,-10){3}{\circle{5}}
\put(25,90){\circle{5}}
\multiput(0,0)(0,40){7}{\line(1,0){40}}
\multiput(0,10)(0,40){7}{\line(1,0){40}}
\multiput(0,20)(0,40){7}{\line(1,0){40}}
\end{picture}%
\begin{picture}(40,260)
\multiput(5,260)(10,-10){3}{\makebox(0,0){$\bullet$}}
\put(5,260){\line(0,-1){132.5}}
\put(15,250){\line(0,-1){132.5}}
\put(25,240){\line(0,-1){112.5}}
% alpha^2
\put(5,200){\circle{5}}
\multiput(15,200)(0,10){2}{\circle{5}}
\multiput(25,200)(0,10){3}{\circle{5}}
% 1
\multiput(5,180)(10,-10){3}{\circle{5}}
% alpha
\multiput(5,140)(0,-10){2}{\circle{5}}
\multiput(15,140)(0,-10){3}{\circle{5}}
\put(25,130){\circle{5}}
\multiput(0,0)(0,40){7}{\line(1,0){40}}
\multiput(0,10)(0,40){7}{\line(1,0){40}}
\multiput(0,20)(0,40){7}{\line(1,0){40}}
\end{picture}%
}
\smallskip
\caption{Alternate version of the encoder shown in figure~\ref{QRSencoder}.\label{QRSencoder2}}
\end{figure}
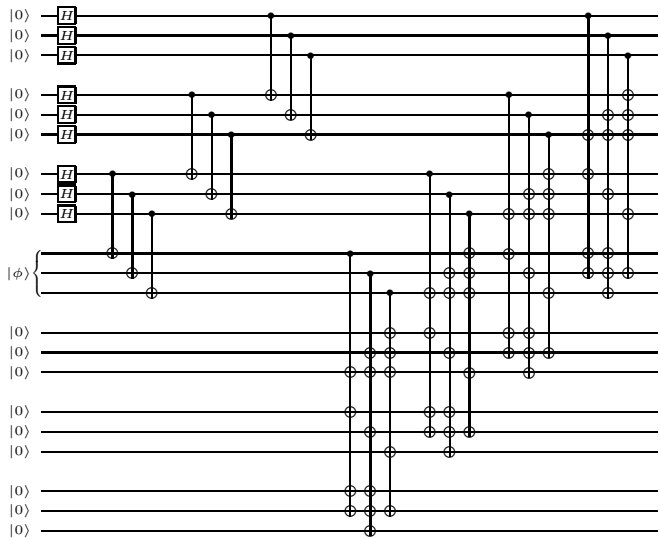

The quantum circuit in figure~\ref{QRSencoder} strictly follows the
concept of cyclic shifting and linear feed-back. Hence it is highly
structured. On the other hand, if shifting cannot be implemented
easily, we can re-shuffle the circuit and simplify it by combining all
shift operations to a permutation of the input (see
figure~\ref{QRSencoder2}).

%==========================================================================
\section{Conclusion}
%==========================================================================
In this paper, we presented new methods for encoding and decoding
cyclic quantum error-correcting codes based on quantum linear shift
registers. They may ease the physical implementation of quantum
computers.

Classically, linear feed-back shift registers are also used to produce
pseudo random sequences for cryptographic purposes. Hence, it is
worthwhile to investigate the cryptographic properties of quantum
states produced by quantum linear feed-back shift registers (QLFSR).

Another application of (classical) linear shift registers is the area
of convolutional codes. Therefore, the quantum version of linear shift
registers might prove useful in the context of quantum convolutional
codes (\noparcite{Cha98}), too.

\begin{acknowledgements}
The authors would like to thank Willi Geiselmann for numerous
stimulating discussions during the process of writing this paper. We
are indebted to Rainer Steinwandt for his critical comments to
preliminary versions of this paper. Part of this work was supported by
{\em Deutsche Forschungsgemeinschaft (DFG), Schwerpunktprogramm
Quanten-Informationsverarbeitung (SPP 1078), Projekt AQUA
(Be~887/13-1)}.
\end{acknowledgements}

\bibliography{QLFSR}
\bibliographystyle{royalsoc}

\label{lastpage}

\end{document}